\documentclass[11pt,a4paper,]{article}
\usepackage{jcappub}
\usepackage{changepage}
\usepackage{amsmath,amssymb}
\usepackage{pmgraph}
\usepackage[normalem]{ulem}
\usepackage[utf8]{inputenc}
\usepackage{wrapfig}
\usepackage{graphicx}
\usepackage{float}
\usepackage{systeme,mathtools}
\usepackage{amsmath,amssymb,upgreek}
\usepackage{xcolor}
\usepackage{soul}
\usepackage{multirow}
\usepackage[toc,page]{appendix}

\usepackage{appendix}
\usepackage{setspace}
\usepackage{subcaption}
\usepackage{hyperref}
\usepackage{import}

\usepackage{siunitx}
\usepackage{mathcomp}
\usepackage{booktabs}
\usepackage{physics}
\usepackage{dsfont}
\usepackage{hyperref}
\usepackage{blindtext}
\newcommand{\overbar}[1]{\mkern 1.5mu\overline{\mkern-1.5mu#1\mkern-1.5mu}\mkern 1.5mu}

\definecolor{darkblue}{RGB}{14,0,185}

\pdfoutput=1
\usepackage{mathrsfs}
\usepackage{epsfig}
\usepackage{rotate}
\usepackage{amsmath}
\usepackage{amssymb}
\usepackage{amsfonts}
\usepackage{enumerate}
\usepackage{afterpage}
\usepackage{color}
\usepackage{comment}

\newcommand{\ud}{\mathrm{d}}
\newcommand{\p}{\partial}
\newcommand{\cH}{\mathcal{H}}
\newcommand{\Per}{\mathcal{P}}
\DeclareUnicodeCharacter{2212}{-}

\newcommand{\pp }{_{\parallel}}
\newcommand{\bb}{\bar}
\newcommand{\ti}{\tilde}
\newcommand{\x}{\mathbf{x}}

\newcommand{\n}{\mathbf{n}}
\newcommand{\at}[2][]{#1|_{#2}}

\newcommand{\D}{\mathcal{D}}

\newcommand{\DL}{\mathscr{D}_L}
\newcommand{\BDL}{\overbar{\mathscr{D}}_{\rm L}}
\newcommand{\BD}{\overbar{\mathcal{D}}}
\newcommand{\WD}{\mathcal{W}_{\D_L}}
\newcommand{\WV}{\mathcal{W}_{\rm V}}

\newcommand{\bn}{\hat{\bf n}}

\def\be{\begin{equation}}
\def\ee{\end{equation}}
\def\bea{\begin{eqnarray}}
\def\eea{\end{eqnarray}}

\title{Gravitational wave  luminosity distance-weighted anisotropies}

\author[a,b]{Andrea Begnoni,}
\author[a,b]{Lorenzo Valbusa Dall'Armi,}
\author[a,b,c]{Daniele Bertacca,}
\author[a,b,c]{Alvise Raccanelli}
\affiliation[a]{Dipartimento di Fisica Galileo Galilei, Università di Padova, I-35131 Padova, Italy}
\affiliation[b]{INFN Sezione di Padova, I-35131 Padova, Italy}
\affiliation[c]{INAF-Osservatorio Astronomico di Padova, Vicolo dell Osservatorio 5, I-35122 Padova, Italy}
\emailAdd{andrea.begnoni@phd.unipd.it}
\emailAdd{lorenzo.valbusadallarmi@unipd.it}
\emailAdd{daniele.bertacca@pd.infn.it}
\emailAdd{alvise.raccanelli@pd.infn.it}

\abstract{Measurements of the luminosity distance of propagating gravitational waves can provide invaluable information on the geometry and content of our Universe. Due to the clustering of cosmic structures, in realistic situations we need to average the luminosity distance of events coming from patches inside a volume. In this work we evaluate, in a gauge-invariant and fully-relativistic treatment, the impact of cosmological perturbations on such averaging process. We find that clustering, lensing and peculiar velocity effects impact estimates for future detectors such as Einstein Telescope, Cosmic Explorer, the Big Bang Observer and DECIGO. The signal-to-noise ratio 
of the angular power spectrum of the average luminosity distance over all the redshift bins is 17 in the case of binary black holes detected by Einstein Telescope and Cosmic Explorer. We also provide fitting formulas for the corrections to the average luminosity distance due to general relativistic effects.}

\arxivnumber{2404.12351}

\begin{document}

\maketitle

\newpage

\section{Introduction}

Since the first detection of Gravitational Waves (GW) from the merger of two black holes~\cite{LIGOScientific:2016aoc}, the network of interferometers LIGO-Virgo-KAGRA (LVK) has confirmed the observation of more than 90 compact binary coalescences (CBC)~\cite{LIGOScientific:2021aug}, including all types of compact object mergers, i.e. binary black holes (BBH), binary neutron stars (BNS) and black hole-neutron star systems (BHNS). In May 2023, the network started the O4 run with an improved sensitivity and it is expected to detect hundreds of CBC per year; furthermore, in the O5 run, planned to start in 2027, around one thousand detections per year are expected~\cite{KAGRA:2013rdx}. A lower instrumental noise does not correspond only to a larger number of resolved events, but also to (on average) higher values of the signal-to-noise ratio (SNR)\cite{ Cutler:1994ys} and more precise estimates of the intrinsic parameters of the binary systems~\cite{Schutz:1986gp,Dalal:2006qt, MacLeod:2007jd, Nissanke:2009kt,Schutz:2011tw}. An accurate determination of the luminosity distance (LD) of these sources 
could be a valuable source of information to reconstruct the expansion history of the Universe. Moreover, given a cosmological model, it can be used to constrain its cosmological parameters~\cite{Maggiore:2007ulw} and as a complementary probe to the electromagnetic sources. 
In particular, we could obtain competitive and independent constraints w.r.t. Type-Ia Supernovae on the Hubble constant~\cite{Chen:2017rfc,Feeney:2018mkj}, which could shed lights on the tension~\cite{Verde:2019ivm,DiValentino:2021izs} between early-~\cite{Planck:2018nkj} and late-~\cite{deJaeger:2020zpb,Riess:2021jrx,SupernovaCosmologyProject:1998vns, Reid:2008nm, Riess:2020fzl, Riess:2023bfx, Sharon:2023ioz, Freedman:2021ahq}  time measurements. In fact, estimates of the LD can be obtained directly from the waveform~\cite{Schutz:1986gp,Finn:1992xs,Cutler:1994ys,Maggiore:2007ulw}, without the need for calibration, earning to these sources the name of ‘‘Standard Sirens"~\cite{Holz:2005df}. Although GW measurements allow to measure LDs with high precision, the redshift of the sources, necessary to evaluate the cosmological parameters given a cosmological model, cannot be estimated by only GW observations. This is due to the full degeneracy between the redshift and the chirp mass of the binaries~\cite{Holz:2005df}, whose contributions to the waveform cannot be disentangled. However, the redshift of the sources, necessary to estimate the cosmological parameters given a cosmological model, cannot be extracted reliably from the waveform, because of its degeneracy with the chirp mass.
The preferred way to solve the issue of the determination of the redshift would require the measurement of an electromagnetic (EM) counterpart (e.g. when a neutron star is involved or when the merger happens in a gas rich environment~\cite{LIGOScientific:2017vwq, Graham:2020gwr}). Anyway, this scenario is too optimistic for most of the sources and for the largest fraction of events it will be not possible to measure the EM counterpart~\cite{Perna:2021rzq}. These sources, when no information on the redshift is present, are called ‘‘Dark Sirens", as opposed to ‘‘Bright Sirens" ~\cite{Borhanian:2020vyr, Finke:2021aom}. In recent years many techniques to provide a competitive estimate of the redshift in the ‘‘Dark Sirens" case have been proposed. 
Possible strategies consist in performing the statistical analysis of the GW sources assuming an astrophysically-motivated distribution of the masses~\cite{Mastrogiovanni:2021wsd, LIGOScientific:2021aug}, exploiting the statistical information on the redshift from galaxy catalogs~\cite{Gair:2022zsa,Gray:2019ksv, Mastrogiovanni:2023emh, Scelfo:2021fqe}, cross-correlating the clustering of GW sources and galaxies~\cite{Oguri:2016dgk,Namikawa:2015prh,Mukherjee:2020hyn,Bera:2020jhx,Raccanelli:2016cud,Cai:2017aea,scelfo2018gw,Bosi:2023amu, Libanore:2023ovr}, making use of external knowledge of the source redshift distribution~\cite{Schutz:1986gp} or of the tidal deformation in the case in which NS are involved~\cite{Messenger:2011gi}.

Since the third-generation interferometers Einstein Telescope (ET)~\cite{Punturo:2010zz,Sathyaprakash:2012jk,Maggiore:2019uih, Cai:2016sby, taylor2012cosmology, Hall:2019xmm} and Cosmic Explorer (CE)~\cite{reitze2019cosmic,Evans:2021gyd}, will detect around $10^5-10^6$ events per year with high significance~\cite{Pieroni:2022bbh,Iacovelli:2022bbs, Muttoni:2023prw, Bellomo:2021mer,Borhanian:2020vyr}, it will be possible to recognize the clustering of the sources due to the anisotropies in the number of host galaxies. 

In order to maximize the information coming from the anisotropies, for each angular patch at a given comoving distance, we weight the LD  w.r.t. the number of sources within this 3D comoving volume. For this reason we must take into account both the fluctuation of LD at the observed position and distance, and the fluctuation of the number density linked at that given patch. These LD measurements will then be influenced by corrections due to LSS (see, e.g.,~\cite{Yoo:2009au,Yoo:2010ni,Bonvin:2011bg, Challinor:2011bk, Jeong:2011as, Bertacca:2012tp}) inhomogeneities through the propagation of GWs from the source to the observer. In other words, we have to consider all  relativistic distortions arising from the observation of the past GW light-cone that could also contribute to our error budget.
These GR corrections affect also the estimates of the LD of the sources as it is well established in the literature. The topic was discussed at linear order in the EM case in~\cite{Pyne:1995iy, Pyne:2003bn,Sasaki:1987ad, Bonvin:2005ps, Hui:2005nm} and in the GW case in~\cite{Bertacca:2017vod, Namikawa:2015prh, Hirata:2010ba, Laguna:2009re}. In this work we investigate the impact of GR corrections on the average LD of a volume containing many sources of GWs. The GR corrections have been computed within the ‘‘Cosmic Ruler" formalism~\cite{Jeong:2011as, Schmidt:2012ne} adopted in~\cite{Bertacca:2017vod} (see also~\cite{Laguna:2009re, Pyne:1995iy}), where the observer’s frame is used as the reference system and the perturbations are computed by looking at the null geodesics of the GWs in a perturbed FLRW metric, which connect the emitter to the observer (e.g. \cite{kristian1966observations,Sasaki:1987ad,sugiura1999anisotropies,Bertacca:2017vod, Bertacca:2017dzm, Cusin:2017fwz, Cusin:2017mjm}). We find that the GR corrections are produced by the perturbations of the LD of the single events computed in~\cite{Bertacca:2017vod}, and by a contribution coming from the anisotropic distribution of the sources (clustering) inside the volume. For a similar procedure\footnote{In the literature an alternative approach has been proposed~\cite{Namikawa:2015prh, Namikawa:2020twf, Fonseca:2023uay, Balaudo:2023klo}. In particular, this perspective is usually referred to as the Luminosity Distance Space}
in the SNIa case see~\cite{Yoo:2019skw,Mitsou:2019ocs}. 

We have implemented the analytical expression of the GR corrections on the weighted LD in a modified version of \texttt{Multi\_Class}~\cite{Bellomo:2020pnw,Bernal:2020pwq}, computing the angular power spectrum of the anisotropies. We estimated also the uncertainties on the average LD due to the shot noise fluctuations in the number of sources and instrumental noise. To account for the instrumental noise we will compute the errors associated to the LD and angular resolution of the single sources by doing a Fisher forecast for BBH and BNS for ET+2CE by using GWFAST~\cite{Iacovelli:2022bbs,Iacovelli:2022mbg} (similar codes are available in the literature e.g. \cite{Dupletsa:2022scg, Borhanian:2020ypi}), while for the shot noise we quantify the number of events consistently with the latest bounds on the populations of BBH and BNS provided by the LVK collaboration~\cite{LIGOScientific:2021djp}. We will show also the case of a BBO~\cite{Cutler:2009qv, Crowder:2005nr}/DECIGO~\cite{Kawamura:2011zz,Kawamura:2020pcg}-like instrument, which are planned to have a much higher sensitivity. As expected, the GR corrections become relevant when the number of events and the sensitivity of the detector are very large.  
We will evaluate the detectability of the angular power spectrum of the GR corrections with third generation interferometers (ET+2CE) and future space-based detectors like BBO/DECIGO. Our findings show that it is possible to measure the contribution to the angular power spectrum of the average LD due to lensing of GWs and density perturbation.

In Section~\ref{motivations} we introduce the weighted LD, in~\ref{The Weighted Luminosity Distance} we perform the analytical calculations of the GR corrections $\DL$, while in~\ref{Angular power spectrum} we  present the covariance matrices of the GR corrections, shot noise and instrumental noise. In~\ref{Results} we illustrate the numerical results and in~\ref{Conclusions} we draw the conclusions and describe the future perspectives and applications of this analysis.

\section{Estimate of the luminosity distance from many sources}
\label{motivations}

\begin{figure}[H]
    \centering
    \includegraphics[width=1\linewidth]{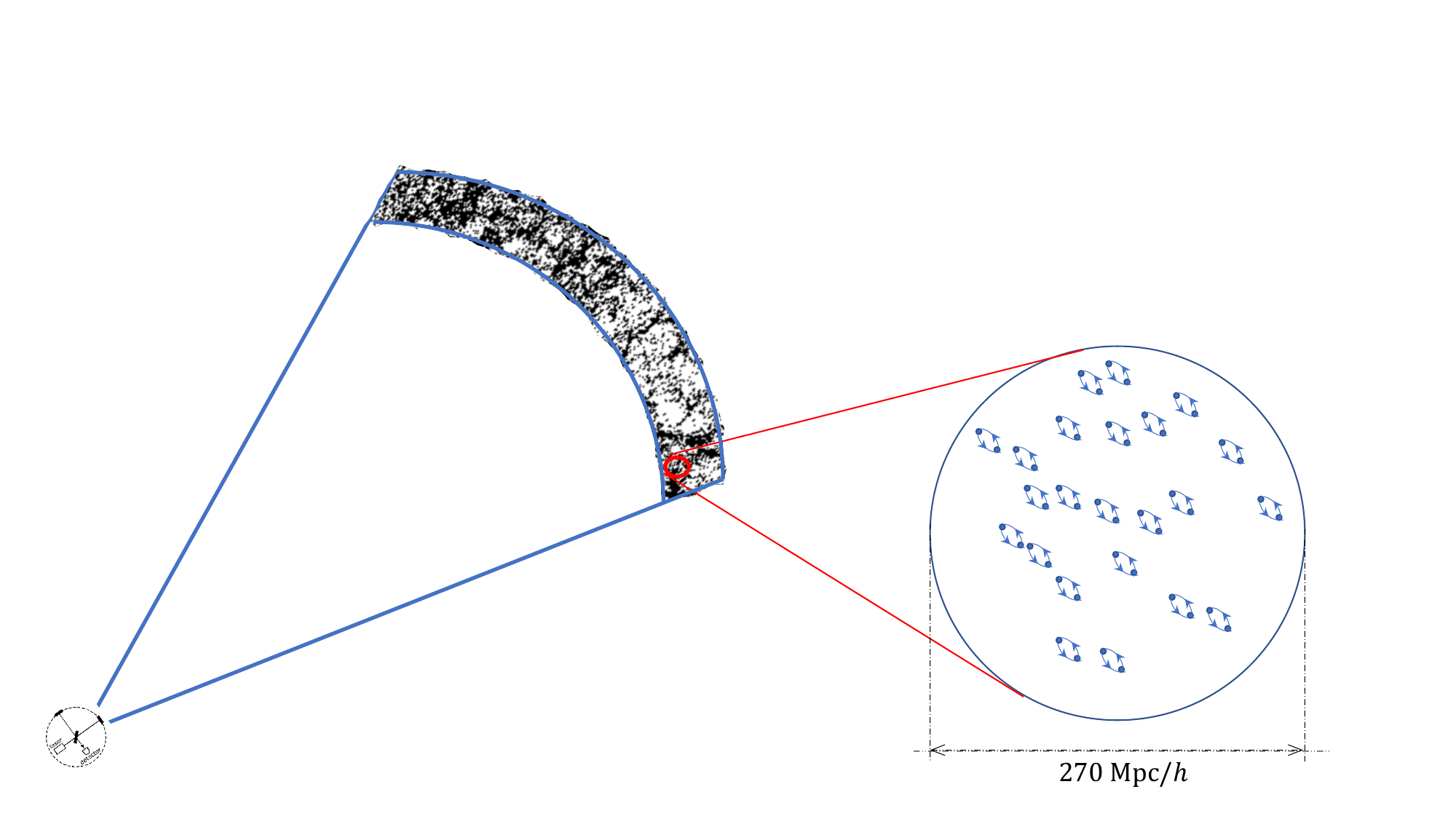}
    \caption{Mock observation of GWs sources in a bin, and zoomed in local distribution.}
    \label{Fig:GWsources}
\end{figure}

In future experiments, a large number of GW sources are expected to be detected, therefore we will have access to a large population of events of which we know (up to a certain precision) the observed LD $\mathcal{D}_{L(i)}$ 
and the position in the sky $\bn_{(i)}$ (i.e. the apparent distance and position). Here the subscript denotes the $(i)$-th GW source that is measured.
The high density of sources implies that we will be able to observe not just the global distribution of the GW events, but also the local density; an example of a mock distribution is shown in Fig.~\ref{Fig:GWsources}.
However, due to instrumental uncertainties and GR corrections, we are unable to have an exact measurement of the observed position $\bn$ and LD $\mathcal{D}_{L}$ of a given resolved GW (unless we can identify a host galaxy with accurate distance and position determination).
In order to exploit the presence of many GW emitters at a given direction and distance, and to reduce the impact of these uncertainties we should consider a weighted average of the LD by using all the sources inside a suitable (small) volume, along the same direction. 
This average we perform, called $\DL$,  is motivated by the fact that the estimate of LD should be more weighted in an over-dense region where more GW events are observed.
The weighted average must be defined in a suitable bin that contains the measured LD, therefore the bin width has to be larger than the uncertainty on the LD and angular resolution of the individual sources. The bin is defined by a window function $W_{\rm b}$ that depends on the angular position and LD of the volume $V_{\rm b}$, where the $b$ indicates the bin. In Eq.~\eqref{eq:def_W_c} we show an example of Gaussian window function for the bin.\\
If the number of objects is low the average tends to a volume average; instead, if the number of the observed GWs is high, at a given direction and distance the estimates of $\DL$ should be intrinsically 
correlated with underlying over-dense matter distribution within $V_{\rm b}$.
Therefore, in the latter case, the clustering distribution could play a non-negligible role in the average over the volume considered. Thus the weighted LDs could be a new cosmological probe, which contain not only the perturbation of the individual LD, but also of the number count.
Let us conclude that a similar prescription has already been suggested in literature~\cite{Mitsou:2019ocs}, for applications see e.g.~\cite{Yoo:2019skw} in the case of SNIa or~\cite{Bernardeau:1995en} in the case of LSS.

In addition, this averaging procedure could be helpful in reducing the computational cost of the joint Bayesian analysis of the many sources which could be detected by future experiments.

In this work, we consider resolved sources, i.e., all the GWs detected with signal-to-noise ratio (SNR) $\varrho$ greater than a certain threshold $\varrho_{\rm thres}$ (which could also depend on the experiment). The population of the resolved events would have an important impact on the anisotropies of the average LD because of their redshift distribution. More details on the SNR threshold and the resolved population are given in Appendix~\ref{app:population_source_and_instrumental_noise}. 
Here, to estimate the average LD of a volume $V_{\rm b}$ we select for simplicity a gaussian window function $W_{\rm b}$, that is peaked at $\left(\D_L^{\rm b},\bn^{\rm b}\right)$, with variance compatible with the experimental uncertainties $\sigma_{\D_L}^{\rm b}$ and $\sigma_{\bn}^{\rm b}$. We picked a gaussian window function to resemble the experimental setup but the same treatment can be applied also to other window functions.  
Then the estimator of the average LD, $\DL$, in the volume $V_{\rm b}$ is defined by 
\begin{equation}
\DL\left(\D_L^{\rm b},\bn^{\rm b}\right) \equiv \frac{1}{N_{\rm GW}^{\rm b}}\sum_{i} W_{\rm b}\left(\D_{L (i)},\bn_{(i)}\right) \D_{L(i)}\, .
    \label{eq:def_DL_average_bin_W_c}
\end{equation}
This equation implicitly takes into account not only the perturbations to the LD at the source location, but also the integral of the perturbations along the line-of-sight, weighted by the window function chosen. Here, as usual, the window function $W_{\rm b}$ is normalized to unity when summing over $\D_{L(i)}$, $\bn_{(i)}$ and we have used the following short notation 
\begin{equation}
    W_{\rm b}\left(\D_{L (i)},\bn_{(i)}\right) \equiv W_{\rm b}\left(\D_{L (i)},\bn_{(i)};\D_L^{\rm b},\bn^{\rm b},\sigma_{\D_L}^{\rm b},\sigma_{\bn}^{\rm b}\right)\, ,
\end{equation}
where $\rm b$ implicitly identifies the volume of the bin considered (e.g. $W_{\rm b_1}$ and $W_{\rm b_2}$ correspond to window functions or distribution of sources associated to different volumes). 

The average of LD at a given $\D_L^{\rm b}$ is indicated with $\langle \DL \rangle_{\hat{\n}}$ and it is computed by averaging the LD 
across all bins in all directions at a given radial shell with a given width $\sigma_{\D_L}^{\rm b}$. Considering the fact that GW experiments are full sky, the expected value of the average LD in all these bins, i.e. $\langle \DL \rangle_{\hat{\n}}$, must be equal to the background value (since all corrections average to zero when integrated in all the solid angle at a fixed shell of $\DL$, which is indicated with $\BDL$. Then we have
\begin{equation}   \label{eq:background_LD_sec2}         
    \left\langle\DL\left(\D_L^{\rm b},\hat{\n}^{\rm b}\right)\right\rangle_{\bn}\equiv \frac{1}{4\pi}\sum_{\hat{\n}^{\rm b}}\DL\left(\D_L^{\rm b},\bn^{\rm b}\right) =
     \BDL\left(\D_L^{\rm b}\right) \, .
\end{equation}
The LD can then be written as the background term plus the GR corrections, the shot noise and the instrumental noise
\begin{equation}
    \DL = \overbar{\mathscr{D}}_L + \Delta\DL + \sigma_{\rm SN} + \sigma_{\rm instr} \, ,
\end{equation}
thus the anisotropic fluctuation at a given  $\mathcal{D}_L^{\rm b}$ can be written in the following way
\begin{equation}
   {\DL\left(\D_L^{\rm b},\bn^{\rm b}\right) \over \overbar{\mathscr{D}}_L\left(\D_L^{\rm b}\right)} -1 = \Delta\ln\DL\left(\D_L^{\rm b},\bn^{\rm b}\right)+\ln \sigma_{\rm SN}\left(\D_L^{\rm b},\bn^{\rm b}\right)+\ln \sigma_{\rm instr}\left(\D_L^{\rm b},\bn^{\rm b}\right)\, , 
   \label{eq:def_bar_DL_discrete}
\end{equation}
where we have defined $\Delta\ln\DL\equiv \Delta\DL/\overbar{\mathscr{D}}_L$, $\ln\sigma_{\rm SN} \equiv \sigma_{\rm SN}/ \overbar{\mathscr{D}}_L$ and $\ln\sigma_{\rm instr} \equiv \sigma_{\rm instr}/ \overbar{\mathscr{D}}_L$. We see that anisotropies can be caused by three terms: GR corrections, shot noise, $\ln \sigma_{\rm SN}$, and instrumental noise, $\ln \sigma_{\rm instr}$. These last two contributions will be discussed in section~\ref{sec:shot_noise}, while in the next sections GR corrections represented by $\Delta\ln\DL$. 

As we will see, the GR corrections contain linear order peculiar velocity (of the emitters of GWs) perturbations, gravitational potential, and gravitational lensing effects generated by 
\begin{itemize}
    \item  the relativistic perturbations of the LD of the single sources and by fluctuations in the number of sources;
    \item projection effects when we compute LD from our past lightcone, i.e., between the GW source and the observer.
\end{itemize}
Finally, the definition of the weighted average LD, given in Eq. \eqref{eq:def_DL_average_bin_W_c}, keeps into account for the density distribution of the sources inside the bin.
It is therefore immediate to see that 
inhomogeneities and anisotropies in the number of GW emitters might affect the error on the estimator. In the next section, we will explicitly compute all these contributions.

\section{The Weighted Luminosity Distance $\DL$}
\label{The Weighted Luminosity Distance}

In this work, we will consider sources of GWs observed in the $\rm Hz-kHz$ frequency range and, consequently, we will assume that their propagation can be described using the \textit{shortwave approximation}~\cite{Isaacson:1968hbi,Isaacson:1968zza}. In this regime, the GWs are described as small ripples which propagate through a slowly-varying perturbed Friedmann-Lemaître-Robertson-Walker background along null geodesics. We identify then with $x^\mu(\chi)$ the comoving coordinates in the physical frame while $\chi$ is the comoving distance. This frame is also called real-space and it is not directly accessible to observations.
The frame over which we perform observations, known as {\it the cosmic laboratory} (CL) frame or observed frame, e.g., see \cite{Bertacca:2014wga, Bertacca:2014hwa}, is instead characterized by a set of coordinates that ‘flattens’ our past gravitational wave-cone, which we indicate with $\overbar{x}^{\mu}$. This implies that an observed merger has the following conformal space-time coordinates
\begin{equation}\begin{split}\label{eq:coo_of_events}
\overbar{x}^{\mu}=(\overbar{\eta},\overbar{\x})=(\eta_0-\overbar{\chi},\overbar{\chi}\hat{\n})\,,
\end{split}\end{equation}
where $\hat{\n}$ 
is the unit vector indicating the location in the sky of the source in this frame, $\eta_0$ is the conformal time of arrival of the GW and $\overbar{\chi}$ is the comoving distance in the CL frame, we will also use $\overbar{\chi}$ as an affine parameter. At the background level it is linked to the LD with the relation $\BD_L = \overbar{\chi} \,(1+z)$.
We will also discuss in Appendix~\ref{app:Cosmic Rulers} the connection at linear order between these two frames.
The coordinates $\overbar{\chi}$, $\BD_L$, $\D_{L,(i)}$, $\hat{\n}_{(i)}$ introduced in Section~\ref{motivations} are evaluated in the CL.  
Here, we derive the weighted LD, let us name it with $\DL$, considering a fully relativistic framework and, in particular, we apply the ‘‘Cosmic Rulers'' formalism introduced in~\cite{Jeong:2011as, Schmidt:2012ne}. In order to compute the real measure of the weighted LD, $\DL$ is computed in the observed frame, taking into account for all the possible effects along the past GW-cone, e.g. cosmological perturbations and inhomogeneities on estimates of the LD of binary mergers.
Let us mention that the non-weighted derivation of the LD for a GW  has already been obtained in~\cite{Bertacca:2017vod} where the authors computed corrections of LD due to velocity, volume, lensing and gravitational potential effects (see also~\cite{Laguna:2009re, Sasaki:1987ad, Bonvin:2005ps}.

Since we perform an average in General Relativity in the physical frame, the weighted LD in a volume\footnote{In the intermediate computations, we will not write explicitly the dependence of $\DL$ on the choice of the bin.} $V_{\rm b}$ can be defined in the following way, by taking the continuum limit of Eq. \eqref{eq:def_DL_average_bin_W_c}
\begin{equation} \begin{split}
    \mathscr{D}_L\left(\D_L^{\rm b},\hat{\n}^{\rm b}\right) &=\frac{\int \ud \x^3 \sqrt{-g(x^{\alpha})}\,n_{\rm GW}^{\rm ph}(x^{\alpha},\varrho > \varrho_{\rm thres})\,\D_L(x^{\alpha})}{\int \ud \x^3\sqrt{-g(x^{\alpha})}\, n_{\rm GW}^{\rm ph}(x^{\alpha},\varrho > \varrho_{\rm thres})} \, .
    \label{eq:def_DL_GR}
\end{split} \end{equation}
While when we move to the CL frame, where we have the observed coordinates we can write
\begin{equation} \begin{split}
    \mathscr{D}_L\left(\D_L^{\rm b},\hat{\n}^{\rm b}\right) &=\frac{\int \ud \overbar{\x}^3 \left|\frac{\p x}{\p \overbar{x}}\right|\sqrt{-g(x^{\alpha}(\overbar{\x}^\beta))}\,n_{\rm GW}^{\rm ph}(x^{\alpha}(\overbar{\x}^\beta),\varrho > \varrho_{\rm thres})\,\D_L(x^{\alpha}(\overbar{\x}^\beta))}{\int \ud \overbar{\x}^3 \left|\frac{\p x}{\p \overbar{x}}\right|\sqrt{-g(x^{\alpha}(\overbar{\x}^\beta))}\, n_{\rm GW}^{\rm ph}(x^{\alpha}(\overbar{\x}^\beta),\varrho > \varrho_{\rm thres})} \, .
\end{split} \end{equation}
Moreover if we introduce a more physical definition of $n^{\rm ph}_{\rm GW}$ we can substitute it with $W_{\rm b}\,n^{\rm ph}_{\rm GW}$ so that we include the window function in the definition of Eq.\eqref{eq:def_DL_GR}, we also avoid writing all the dependencies
\begin{equation} \begin{split}
    \mathscr{D}_L\left(\D_L^{\rm b},\hat{\n}^{\rm b}\right) &=\frac{\int \ud \overbar{\x}^3 \left|\frac{\p x}{\p \overbar{x}}\right| \sqrt{-g}\,  W_{\rm b}\,n_{\rm GW}^{\rm ph}\,\D_L}{\int \ud \overbar{\x}^3 \left|\frac{\p x}{\p \overbar{x}}\right|\sqrt{-g}\,  W_{\rm b}\, n_{\rm GW}^{\rm ph}} \, ,
\end{split} \end{equation}
where we stressed that the window function is defined in the CL coordinates.
As was highlighted in Section~\ref{motivations}, the window function of the considered bin has been directly normalized to unity in the CL frame (i.e. when we directly use observed (averaged) coordinates)
\begin{equation}\begin{split}
    \int \ud \overbar{\x}^3 \, \, W_{\rm b}(\overbar{x}^\alpha) = 1 \, ,
\end{split}\end{equation}
while the denominator of Eq.~(\ref{eq:def_DL_GR}) represents the total number of GW sources in the volume $V_{\rm b}$, 
\begin{equation}\begin{split}
    N_{\rm GW}^{\rm b}\left(\D_L^{\rm b},\hat{\n}^{\rm b}\right) \equiv \int \ud \x^3 \sqrt{-g}\,  W_{\rm b}\,  n^{\rm ph}_{\rm GW}\, .
    \label{def:N_gw_bin}
\end{split}\end{equation}
Moreover $g$ is the trace of the metric $g_{\mu\nu}$ in the physical frame. Details on the metric and geodesics of the GWs have been provided in Appendix~\ref{app:Cosmic Rulers}. To improve  readability, when we write a quantity in the CL frame we avoid specify the coordinates, implying that $\overbar{n}_{\rm GW}(\overbar{x}^{\alpha})=\overbar{n}_{\rm GW}$, while for the physical frame we have $n_{\rm GW}(x^{\alpha})=n_{\rm GW}$. Note that the average LD defined by Eq. \eqref{eq:def_DL_GR} is not the geometrical center of the bin, called here $\D_L^{\rm b}$. This is due to the fact that the former represents the average of the LD weighted by the window function of the bin and by the number of sources, while the latter is defined as the maximum of the window function.
We underlined the difference between $\DL$ and $\D_L^{\rm b}$, because the geometrical center of the bin is not an observable, but a quantity chosen depending on the analysis we are interested in. When the distribution of the sources in the sky is uniform, then $\DL$ coincides with $\D_L^{\rm b}$. The number of sources in the bin, defined in Eq.~\eqref{def:N_gw_bin}, contains contaminations from the GR corrections, shot noise and instrumental noise. Similarly to the procedure done in \eqref{eq:background_LD_sec2} we can obtain the background number density, defined as
\begin{equation}\begin{split}
    \overbar{N}^{\rm b}_{\rm GW}=\int \ud \overbar{\x}^3 \,W_{\rm b}\,\overbar{n}_{\rm GW}\, ,
\end{split}\end{equation}
where the physical number density of the GW sources has been replaced by  
\begin{equation}\begin{split}
    \overbar{n}_{\rm GW}^{\rm ph} = \frac{\overbar{n}_{\rm GW}}{\overbar{a}^3}\,,
\label{eq:def_n_gw}
\end{split}\end{equation}
where $\overbar{a}$ is the scale factor of the FLRW Universe at $\overbar{\eta}$ in the observed frame.
The background density of events is obtained by averaging over all the sky the number of events at a given distance from the observer. While $\overbar{\mathscr{D}}_L$, can be written in the continuous case as
\begin{equation}\begin{split} 
    \label{DL background}
    \overbar{\mathscr{D}}_L\equiv\frac{\int \ud\overbar{\x}^3\, W_{\rm b}\,\overbar{n}_{\rm GW}\,\BD_L}{\int \ud\overbar{\x}^3\, W_{\rm b}\,\overbar{n}_{\rm GW}}=\frac{1}{\overbar{N}_{\rm GW}^{\rm b}}\int \ud\overbar{\x}^3 W_{\rm b}\,\overbar{n}_{\rm GW}\,\BD_L\, ,
\end{split}\end{equation} 
where $\BD_L$ is the background LD, obtained by averaging inside a spherical bin. 
By looking at Eq.~\eqref{eq:def_DL_GR}, it is clear that the GR corrections impact the average LD both through fluctuations in $n_{\rm GW}^{\rm ph}$ and $\D_L$. The GR corrections on the number density of events have been computed for instance in~\cite{Yoo:2009au,Yoo:2010ni,Bonvin:2011bg, Challinor:2011bk,Bertacca:2012tp}, while for the LD of a single event in~\cite{Bertacca:2017vod}. Following the notation used in~\cite{DiDio:2013bqa}, we therefore characterize these corrections by introducing the source functions $\Delta_{\rm GW}$ and $\Delta\ln\D_L$. The explicit expressions of these source functions have been given in Appendix~\ref{app:Relativistic contributions terms}. The computation of the perturbation of the average LD gives
\begin{equation} \begin{split}
    \Delta\ln\DL= \int\ud\overbar{\x}^3\,\left[ \mathcal{W}_{\D_L}\Delta\ln\D_L+\left(\WD-\WV\right)\Delta_{\rm GW}\right]\, .
    \label{eq:decomposition_W_DL}
\end{split} \end{equation}
where we have introduced the window functions
\begin{equation} \begin{split}
    \mathcal{W}_V\left(\overbar{x}^i\right)\equiv &\frac{W_{\rm b} \, \overbar{n}_{\rm GW}}{\int \ud\overbar{\x}^3\,W_{\rm b} \, \overbar{n}_{\rm GW}}=\frac{W_{\rm b}}{\overbar{N}^{\rm b}_{\rm GW}}\,\overbar{n}_{\rm GW}\,, \\
    \mathcal{W}_{\D_L} \left(\overbar{x}^i\right)  \equiv&\frac{W_{\rm b}\,\BD_L\,\overbar{n}_{\rm GW}}{\int \ud\overbar{\x}^3\, W_{\rm b}  \, \BD_L\, \overbar{n}_{\rm GW}}\, .
    \label{def:window_functions_V_DL}
\end{split} \end{equation}


The overall correction $\Delta\ln\DL$ is therefore the sum of two terms. The first one is the perturbation of the LD computed, e.g. in~\cite{Bertacca:2017vod}, averaged, directly in the CL frame, over many sources in the bin, while the second term represent the contribution of the clustering of sources due to the weighting averaging procedure. In this work, we compute and evaluate the cosmological signal of $\Delta \ln \DL$ for the GW case. In Appendix~\ref{app:Relativistic contributions terms}, we compute the linear corrections in a generic gauge, while here we write the result, for simplicity, in the Poisson gauge~\cite{Bardeen:1980kt}, defined by the linear line element 
\begin{equation}\begin{split}\label{metric poisson}
\ud s^2=a^2(\eta)[-(1+2\Phi)\ud\eta^2+(1-2\Psi)\delta_{ij}\ud x^i\ud x^j] \, . 
\end{split}\end{equation}
In $\Lambda$CDM, after the equality between matter and radiation, the anisotropic stress of neutrinos is negligible and it is possible to assume $\Psi=\Phi$ at linear order. The perturbation of the average LD in this gauge is then
\begin{equation}
    \begin{split}
    \label{eq:total formula pois}
    \Delta\ln\DL=&\,\, \int \ud\overbar{\x}^3\Biggl\{\WD\biggl\{\left (-\frac{1}{\overbar{\chi}\cH}+1\right)v\pp+\left (\frac{1}{\overbar{\chi}\cH}-2\right)\Phi- \kappa_{\rm p}-\frac{T_{\rm p}}{\overbar{\chi}}+2\left (-\frac{1}{\overbar{\chi}\cH}+1\right)I_{\rm p}\\
    &+\frac{1}{\overbar{\chi}}\delta x^0_o+\left (\frac{1}{\overbar{\chi}\cH}\right)v_{\|,o}+\left (-\frac{1}{\overbar{\chi}\cH}+1\right)(\delta a_o+\Phi_o)\biggl\} \\
    &+\left(\WD-\WV\right)\biggl\{b_{\rm GW}\delta_m^{\rm SC}+\left [-\frac{(5s-2)}{\overbar{\chi}\cH}+10s-1-b_{\rm e}+\frac{\cH'}{\cH^2}\right]\Phi\\
    &+\left [\frac{(5s-2)}{\overbar{\chi}\cH}-5s+b_{\rm e}-\frac{\cH'}{\cH^2}\right]v\pp-\frac{1}{\cH}\p\pp v\pp-\left(b_{\rm e}-3\right)\cH v\\
    & +\frac{1}{\cH}\Phi'+\left [\frac{(10s-4)}{\overbar{\chi}\cH}-10s+2b_{\rm e}-2\frac{\cH'}{\cH^2}\right]I_{\rm p}\\
    &+(5s-2)\left ( \kappa_{\rm p}+\frac{T_{\rm p}}{\overbar{\chi}}\right)-\frac{(5s-2)}{\overbar{\chi}}\delta x^0_o+\left [-\frac{(5s-2)}{\overbar{\chi}\cH}+2-b_{\rm e}+\frac{\cH'}{\cH^2}\right]v_{\|,o}\\
    &+\left [\frac{(5s-2)}{\overbar{\chi}\cH}-5s+b_{\rm e}-\frac{\cH'}{\cH^2}\right](\delta a_o+\Phi_o)\biggl\}\Biggl\}\, ,
    \end{split} 
\end{equation}
with $\cH=\ud \overbar{a} / (\overbar{a}\ud \eta)$ the comoving Hubble rate, $v^i$ is the gauge invariant velocity perturbation, $v\pp = v^in_i$ the component of the peculiar velocity of the sources along the line-of-sight (here we impose that the peculiar velocity is the same as that of the host galaxy, this is reasonable if we consider anisotropies on very large scales), $v$ is the velocity potential, defined as $v_i=\p_i v$, where any irrotational contribution to the velocity is negligible~\cite{Bernardeau:2001qr}. 
Moreover $\delta_m^{\rm SC}$ is the density perturbation in the synchronous-comoving gauge (e.g.~\cite{Bertacca:2014hwa}), it is connected to the density perturbation in the Poisson gauge by Eq.~\eqref{eq:delta_m_Poisson_SC}; the bias $b_{\rm GW}$ is evaluated in the SC gauge.  The suffix $o$ indicates terms evaluated at the observer and it consists of a monopole, proportional to $\Phi_o$ ($\delta a_o$ and $\delta x^0_o$ depend both on $\Phi_o$), and a dipole, proportional to our peculiar motion $v_{\parallel,o}$. We have also written the Integrated Sachs-Wolfe effect~\cite{Sachs:1967er} and the Shapiro time delay~\cite{Shapiro:1964uw} in the compact forms
\begin{equation} \begin{split}
    I_{\rm p}\equiv-\int_0^{\overbar{\chi}}\ud\ti\chi\,\Phi'\,,\quad \quad T_{\rm p}=-2\int^{\overbar{\chi}}_0 \ud\ti\chi\,\Phi\,,
\end{split} \end{equation}
where the suffix $\rm p$ indicates that they are evaluated in the Poisson gauge, also the prime indicates the derivative w.r.t. the comoving time $\eta$. Furthermore $\hat{\kappa}_{\rm p}$ is the weak lensing convergence, defined as
\begin{equation} \begin{split}
        \hat{\kappa}_{\rm p}
    &=\frac{\delta x_{\|,o}}{\overbar{\chi}}-v_{\|,o}+\int_0^{\overbar{\chi}}\ud\ti\chi(\overbar{\chi}-\ti\chi)\frac{\ti\chi}{\overbar{\chi}}\ti\nabla_{\perp}^2\,\Phi=\kappa_{\rm p,o}+\kappa_{\rm p}\,,
\end{split} \end{equation}
with $\kappa_{\rm p}$ the integrated terms and with $\kappa_{\rm p,o}$ the terms at the observer. Following~\cite{scelfo2018gw, Scelfo:2020jyw, Zazzera:2023kjg}, we have defined the magnification bias $s$ of the GWs as
\begin{equation}\begin{split}
    \label{magnification bias scelfo}
    s\left(\BD_L\right) \equiv - \frac{1}{5}\frac{\ud\ln \overbar{n}_{\rm GW}(\BD_L,\varrho > \varrho_{\rm thres})} {\ud \ln\varrho}\at[\bigg]{\varrho=\varrho_{\rm thres}}\, ,
\end{split}\end{equation}
which enters in Eq.\eqref{eq:total formula pois} and represents how the change in the number of observed GW sources is affected by the slope of the faint-end SNR distribution.
The magnification bias $s$ depends on the choice of the threshold and on the population of the sources and it is affected by the sources detected with a SNR close to $\varrho_{\rm thres}$. In particular, the threshold $\rho_{\rm thres}$ to claim a detection is constant throughout the LD, while the SNR of a source decreases going to higher LD. This affects obviously the detectability of the source and consequently the magnification bias. In the next section we will describe the procedure used to evaluate it. The evolution bias, on the other hand, encodes the information about the anisotropies due to the creation of new sources~\cite{Bertacca:2014hwa} and has been defined by
\begin{equation}
    \begin{split}
        b_{\rm e}\equiv \frac{\p\, \overbar{n}_{\rm GW}}{\p\ln\overbar{a}}+3\, .
    \label{def:evolution_bias}
    \end{split}
\end{equation}
Following~\cite{Jeong:2011as, Schmidt:2012ne}, in the previous equations we have also decomposed spatial vectors and tensors into their irrotational and solenoidal components,
\begin{equation}\begin{split}
    A\pp=n^in^j A_{ij}\,,\qquad B^i_{\perp}=\Per^{ij}B_j=(\delta^{ij}-n^in^j)B_j\, ,
\end{split}\end{equation}
while for the derivative we have used the definition $\nabla_{\perp}^2 \equiv \Per^{ij}\p_j\left(\Per_{ik}\p^k\right)$. Moreover
$\delta x_o^0$ and $\delta a_0$ are the perturbation of the time coordinate and of the scale factor at the observer and their expressions have been given in Eqs.~\eqref{eq:pert_at_the_observer}.
Eq.~\eqref{eq:total formula pois} is the main theoretical result of this work and it contains all the physical information we are interested in. In the following sections, we will compute the angular power spectrum of this correction, quantifying its impact on the estimate of the average LD with future experiments. 

\section{Angular power spectrum}\label{Angular power spectrum}
Since with future third-generation detectors we will detect many sources, it is useful to decompose the observed field in spherical harmonics. This would help us understand which contributions dominate at a given angular scale. We focus on the angular power spectrum of the GR corrections since this will give us cosmological information about evolution and anisotropies.
To do so we start from Eq.\eqref{eq:decomposition_W_DL}, which can be rewritten as 
\begin{equation}
    \begin{split}
        \Delta\ln\DL = \int \ud\overbar{\x}^3\sum_i W_i\, \Delta_i\,,
    \end{split}
\end{equation}
where
\begin{equation}
    \begin{split}
        \left.W_{i}\Delta_{i}\right|_{i=0} &= \WD\Delta\ln\D_L\\
        \left.W_{i}\Delta_{i}\right|_{i=1} &=\left(\WD-\WV\right)\Delta_{\rm GW}\,.
    \end{split}
\end{equation}
Now we can follow the standard notation (e.g.  \cite{DiDio:2013bqa}) by writing
\begin{equation}
    \begin{split}
        \Delta_i(\mathbf{n}, \D_L)=\sum_{\ell m} a_{\ell m}(\D_L) Y_{\ell m}(\mathbf{n}),\quad {\rm with} \quad a_{\ell m}(\D_L)=\int d \Omega_{\mathbf{n}} Y_{\ell m}^*(\mathbf{n}) \Delta_i(\mathbf{n}, \D_L)\,,
    \end{split}
\end{equation}
and then, moving to Fourier space we can introduce
\begin{equation}
    \begin{split}
       \left[\Delta\ln\DL\right]_\ell(\D_L^{\rm b},k)=\int \ud \BD_L \sum_i W_i \Delta_{i\,\ell}(k)\,,
    \end{split}
\end{equation}
where the integral over the angles is performed inside the definitions of $\Delta_{i\,\ell}$. Moreover 
\begin{equation}\label{eq:delta_GW_as_sum_of_terms}
    \begin{split}
        \left[\Delta\ln\D_L\right]_\ell=& \left[\Delta\ln\D_L\right]_\ell^{\rm vel}+\left[\Delta\ln\D_L\right]_\ell^{\rm lens}+\left[\Delta\ln\D_L\right]_\ell^{\rm GR}\,,\\
        \Delta_{\rm GW\,\ell}=& \Delta_\ell^{\rm dens}+\Delta_\ell^{\rm vel}+\Delta_\ell^{\rm lens}+\Delta_\ell^{\rm GR}\,,
    \end{split}
\end{equation}
with the definition of each term given in Appendix \ref{app:Relativistic contributions terms}.
Then the angular power spectrum can be written as
\begin{equation}
    \label{Cl general}
    C^{\DL}_{\ell}\left(\D_{L,i}^{\rm b},\D_{L,j}^{\rm b}\right)=4\pi\int\frac{\ud k}{k}P(k)\left[\Delta\ln\DL\right]_\ell\left(\D_{L,i}^{\rm b},k\right)\left[\Delta\ln\DL\right]_\ell^*\left(\D_{L,j}^{\rm b},k\right) ,
\end{equation}
where the star identifies the complex conjugate, while $\D_{L,i}^{\rm b}$ the LD of the $i$-th bin and $P(k)$ is the matter power spectrum. From now on let us simplify the notation, writing $C_{\ell}\equiv C_\ell\left(\D_{L,i}^{\rm b},\D_{L,i}^{\rm b}\right)$, when we consider the angular power spectrum in the same bin.

\subsection{Binning the population of GW sources}
\label{sec:Binning the population of GW sources}

\begin{figure}[t!]
    \centering
    \includegraphics[scale=0.6]{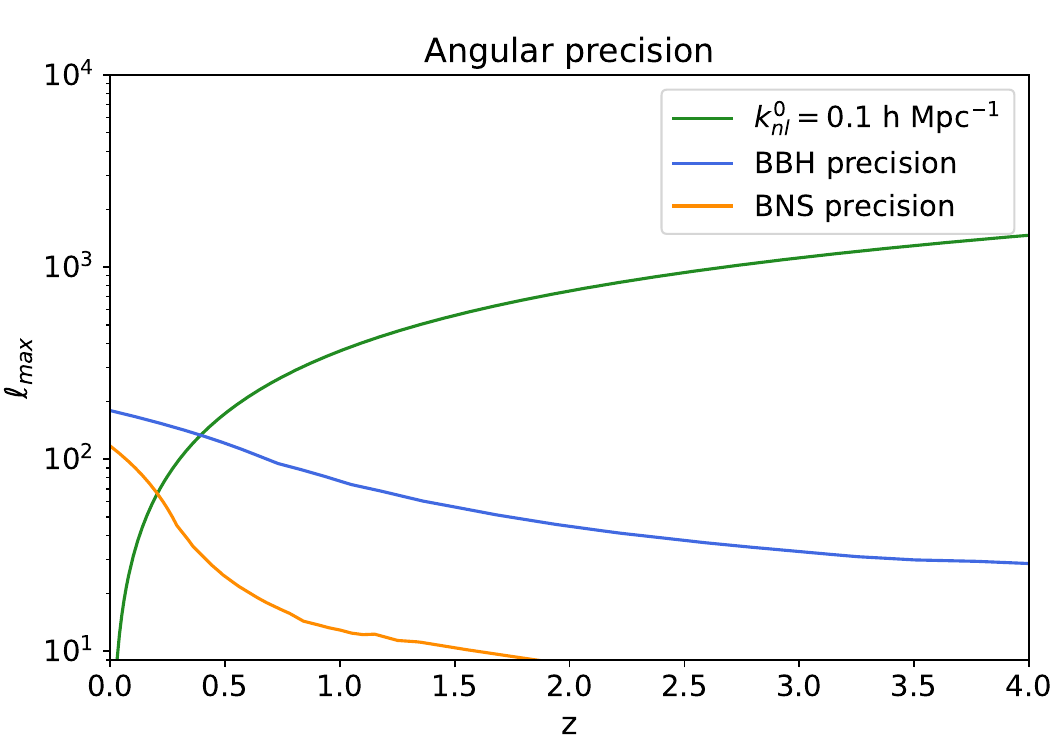}
    \caption{Plot of the comparison between the angular precision expected from ET+2CE, translated in multipoles, and the non-linear scales according to the values of $k^0_{nl}=0.1\,\rm h\,Mpc^{-1}$, BBH are dominated by non-linearities for $z\lesssim 0.4 - 0.5$, while BNS for $z \lesssim 0.2-0.3$. 
    Therefore, in our analysis, we consider multipoles up to the non-linear scale at low redshift, while, as redshift grows, the limiting factor becomes the instrumental precision.}
    \label{fig:non_linear_scale}
\end{figure}

The source functions of the angular power spectra introduced in Eq.~\eqref{Cl general} depend on the window function of the bin $W_{\rm b}$. We consider for simplicity $W_{\rm b}$ 
as the product of two Gaussians in the LD and in the sky position, centered at $\left(\D^{\rm b}_L,\hat{\n}^{\rm b}\right)$, 
\begin{equation}
    \begin{split}
    W_{\rm b}\left(\D_L,\hat{\n}\right) =& \frac{1}{\sqrt{2\pi}\sigma_{\D_L}^{\rm b}}{\rm exp}\left[-\left(\frac{\D_L-\D_L^{\rm b}}{\sqrt{2}\sigma_{\D_L}^{\rm b}}\right)^2\right]\frac{1}{\sqrt{2\pi}\sigma_{\hat{\n}}^{\rm b}}{\rm exp}\left[-\left(\frac{\left|\hat{\n}-\hat{\n}^{\rm b}\right|}{\sqrt{2}\sigma_{\hat{\n}}^{\rm b}}\right)^2\right] \, .
    \label{eq:def_W_c}
    \end{split}
\end{equation}
From now on we will avoid writing the overbar to indicate $\BD_L$, unless necessary, to simplify the notation. The parameters $\D_L^{\rm b}$, $\hat{\n}^{\rm b}$ identify the radial and angular center of the bin and $\sigma_{D_L}^{\rm b}$, $\sigma_{\hat{\n}}^{\rm b}$ the corresponding widths. 
The lower bound of the widths of the bin is set by the average over the population of the instrumental noise of LD and the angular resolution of each source, in this work identified by $\sigma_{\D_L}(\D_L)$ and $\sigma_{\hat{\n}}(\D_L)$.
The instrumental uncertainties have been obtained from a Fisher matrix analysis on the intrinsic parameters of the binaries of a catalog of BBH and BNS generated according to the latest constrains from LVK~\cite{LIGOScientific:2021djp}. We have performed the analysis with the code GWFAST~\cite{Iacovelli:2022bbs, Iacovelli:2022mbg}, considering ET+2CE with $\varrho_{\rm thres}=12$. Additional details on the detected population of events and their associated errors have been given in Appendix~\ref{app:population_source_and_instrumental_noise}.
We decided then to work with a sub-sample of the full detected events, where the selection has been based on their angular resolution.
In this way it is possible to select events with better angular resolution and study the anisotropies at higher multipoles.  
Although $\ell_{\rm max}$ increases when $\sigma_{\hat{\n}}$ decreases, a lower number of events considered would augment the impact of the shot noise, which is roughly inversely proportional to the number of events in the bin (see Section~\ref{sec:shot_noise} for more details). Therefore we choose events with $\sigma_{\hat{\n}}<20\, \rm deg^2$ and $\sigma_{\hat{\n}}<40\, \rm deg^2$ for BBH and BNS to maximize the SNR of the GR corrections on $\DL$ defined in Eq.~\eqref{eq:SNR_pseudo_Cl_total}. For more details on the choice of this sub-sample we refer to Appendix~\ref{app:population_source_and_instrumental_noise}, while a complete definition of the bins used is given in Appendix \ref{sec:Binning}. In Figure~\ref{fig:rho_threshold_quantities} we plot the mean uncertainties on the LD and angular resolution as a function of LD 
(the top x-axis) and redshift (the bottom x-axis) for the two populations of BBH and BNS. 
In the same Figure 
, we also plotted the number of events in one year of observations and the magnification bias for ET+2CE, as a function of the redshift and LD. In the case of BBO/DECIGO (the dotted curves in the top left panel of Figure~\ref{fig:rho_threshold_quantities}) we assume that all of the sources will be detected with high enough SNR~\cite{Cutler:2009qv}, therefore we consider the full BBH and BNS populations and we set the magnification bias to zero, since all the sources are detected with SNR much larger than the threshold. 
\begin{figure}
    \centering
    \includegraphics[width=\linewidth]{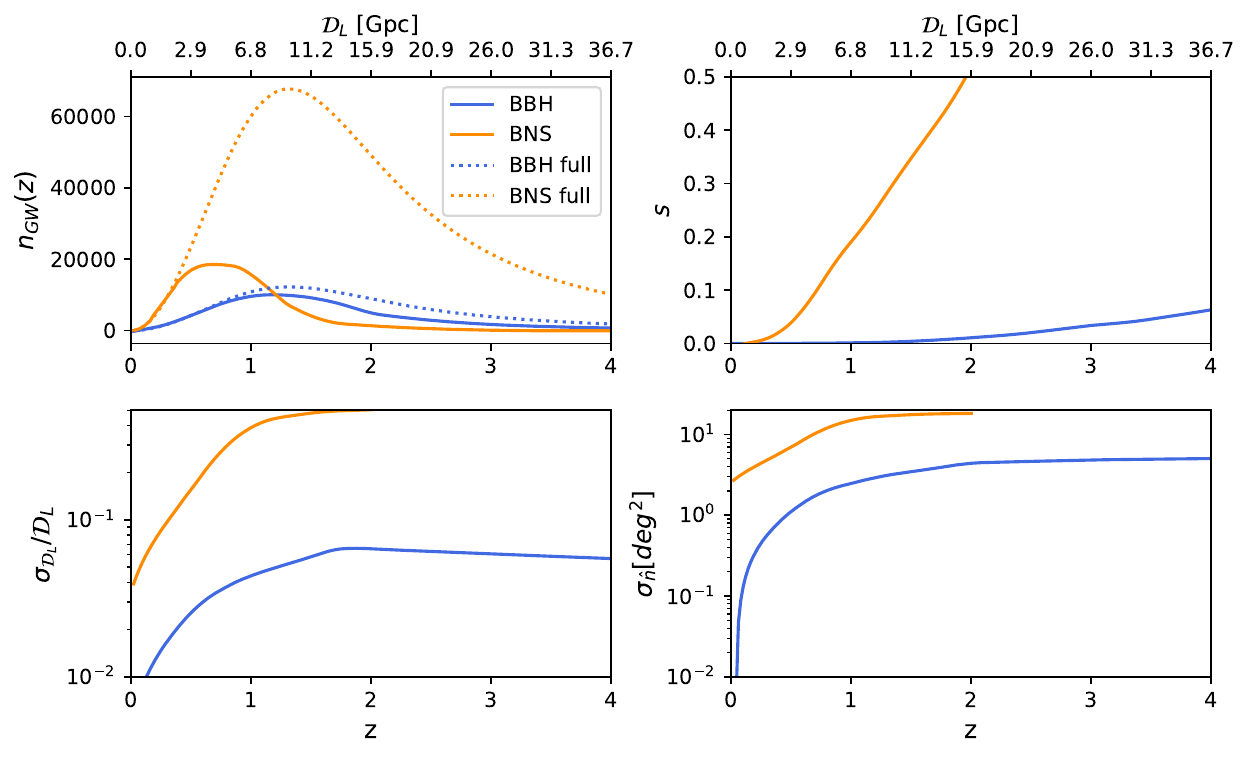}
   \caption{Top left: plot of the number of resolved sources by ET+2CE as a function of the redshift and LD for BBH and BNS in one year of observations, for the sub-sample used in this work (solid lines), and full populations (dashed). Top right: plot of the magnification bias for BBH and BNS for ET+2CE. Bottom panels: relative errors on the LD (left) and sky-localization (right) for our sub-population. 
   } 
    \label{fig:rho_threshold_quantities}
\end{figure} 
The smallest observable scales of the anisotropies of $\DL$ depend on error associated to angular resolution, $\sigma_{\hat{\n}}$.
This information is accounted for in the exponential in $\hat{\n}$ in Eq.~\eqref{eq:def_W_c}, which goes to zero when $\left|\hat{\n}-\hat{\n}^b\right|\ll \sigma_{\hat{\n}}^{\rm b}$, forbidding the access to scales smaller than the angular resolution of the instrument. 
To take into account for this, we decide to use the strategy adopted in~\cite{Mukherjee:2019wcg,Libanore:2020fim,Libanore:2021jqv}, in which a Gaussian beam profile, which exponentially suppresses the anisotropies at multipoles smaller than $\ell^{\rm res}_{\rm max}$, is introduced. The correspondence between $\sigma_{\hat{\n}}$ and the $\ell_{\rm max}$ used in the analysis has been shown in Figure~\ref{fig:non_linear_scale}. 
More specifically, the angular power spectrum of the GR corrections, computed in Eqs.~\eqref{Cl general}, is convolved in the likelihood with the beam 
\begin{equation}
    B_\ell\left(\D_{L}^{\rm b}\right) \equiv {\rm exp}\left[-\frac{\ell(\ell+1)}{8\ln 2}\sigma^2_{\hat{\n}}\left(\D_{L}^{\rm b}\right)\right]\, .
    \label{def:cut_ell_max_instr}
\end{equation}

We use the same prescription of~\cite{Libanore:2020fim,Libanore:2021jqv}, where the maximum multipole $\ell^{\rm nl}_{\rm max}$ at which we computed the angular power spectrum is defined by
\begin{equation}
    \ell^{\rm nl}_{\rm max}\left(\D_{L}^{\rm b}\right)\equiv k_0^{\rm nl}\left[1+z^{\frac{2}{2+n_s}}\left(\D_{L}^{\rm b}\right)\right]\chi\left(\D_{L}^{\rm b}\right)\, .
\end{equation} 
where $k^{\rm nl}_0=0.1 \,\rm h\,Mpc^{-1}$ and
$n_s$ is the spectral tilt of the scalar power spectrum.
The previous formula links the non-linear scale today to the one at redshift $z$ and then convert that scale into multipoles. Since we want to constrain just linear cosmological perturbation theory, our analysis of the anisotropies of $\DL$ will stop at $\ell_{\rm max}^{\rm nl}$.
We limit the analysis for BNS at $z=2$, because of the low number of detected sources at larger redshifts.
In the case of BBO/DECIGO we consider the precisions from \cite{Holz:2005df,camera2013beyond}. The larger sensitivities of BBO/DECIGO allows to reach for both BBH and BNS an improvement of at least one order of magnitude in the angular precision and the LD w.r.t. ET+2CE. These interferomenters could indeed measure the LD with a relative error of the order of sub-percent and the sky position up to arcsec. 
In this work we choose equally spaced gaussian bins in $z$, while their width is determined by experimental uncertainties as described in this section. We also verified that our results are robust against different binning strategies, while for more information on the binning we refer to Appendix \ref{sec:Binning}.


\subsection{Intrinsic Anisotropies}
\label{Intrinsic Anisotropies}

The intrinsic anisotropies of the weighted LD, generated by cosmological perturbations, are due to density perturbations of matter, 
the velocity perturbation between the sources and the observer (represented by the Doppler shift and the redshift space distortion), the lensing term and GR effects due to metric perturbations. According to Eq.~\eqref{Cl general}, the angular power spectrum can be computed as the convolution of two source functions $\left[\Delta\ln\DL\right]_\ell$ with the scalar power spectrum. 
To compute numerically the angular power spectrum, we used a modified version of \texttt{Multi\_CLASS}~\cite{Bellomo:2020pnw, Bernal:2020pwq}. Following the notation of \texttt{CLASSgal}~\cite{DiDio:2013bqa}, we distinguish among four contributions to the anisotropies of $\Delta\ln\DL$: density, velocity, gravitational potentials and lensing terms. In Eq.~\eqref{eq:source_functions_Weighted_DL}, we have provided the source functions for all these terms. The density contribution is due to the overdensity in the number of sources and it is connected to the underlying matter distribution. The velocity term accounts for all the relativistic effects associated to the peculiar motion of the sources, while the gravitational contribution contains perturbations generated by the scalar perturbations of the metric, such as the Integrated Sachs-Wolfe (ISW) and are indicated under the term GR terms. The lensing perturbation consists in the magnification of the GWs by gravitational lenses. 

\begin{figure}[t!]
    \centering
    \includegraphics[width=\linewidth]{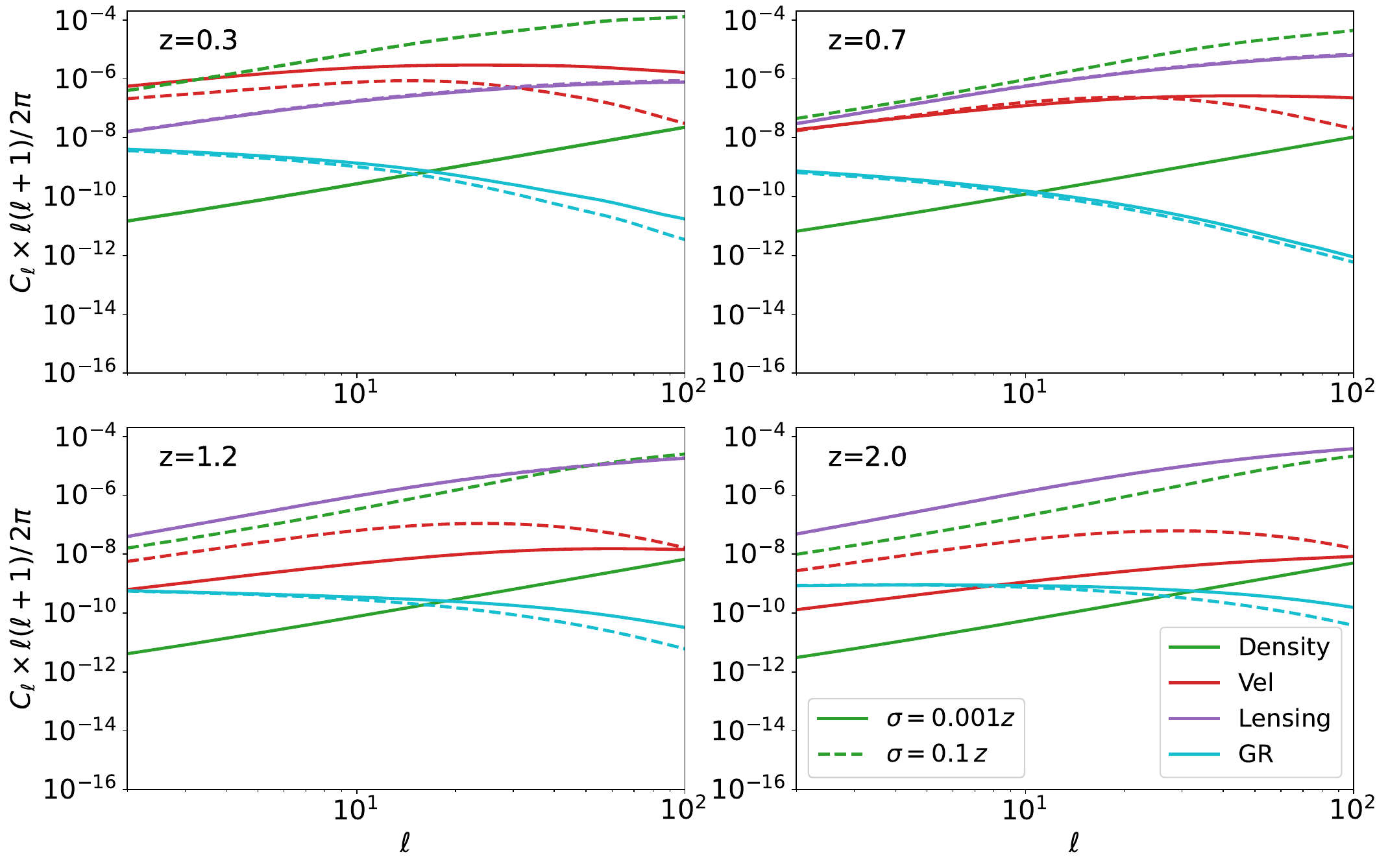}
    \caption{Plot of the density, velocity, GR and lensing contributions to the angular power spectrum of $\Delta\ln\DL$, for bin centered at redshift $0.3$ (top left), $0.7$ (top right), $1.2$ (bottom left) and $2.0$ (bottom right). Solid lines are for $\sigma=0.001z$, dashed ones $\sigma=0.1z$.}
    \label{fig:Cl_GR_corrections}
\end{figure}

In Figure~\ref{fig:Cl_GR_corrections} we plot the angular power spectra, in the same redshift bin, of contributions
to the GR corrections for four different bins (i.e. $z=0.3,\,0.7,\,1.2,\,2.0$) and two choices of 
$\sigma_{\D_L}^{\rm b}$ (i.e. $\sigma_{\D_L}^{\rm b}=0.1z,\,0.001z$). In order to make an easier comparison with other angular power spectra computed in  literature, we write the mean and the width of the bins in terms of the redshift and not of the LD. 
As expected, for small widths of the bins the result converges to the findings of~\cite{Bertacca:2017vod} and the density contribution goes to zero. On the other hand, for larger bin sizes, such as e.g.,~$\sigma_z = 0.1\,z$, the density contribution is larger at low redshifts and it is dominant for $z\lesssim0.7$, as can be seen in Fig.~\ref{fig:Cl_GR_corrections}. The lensing contribution is the dominant one at high redshifts, so from $z\gtrsim1.2$. Since the lensing is an integrated effect, it is the least affected by the choice of the bin width. In fact its value is close to the one calculated in \cite{Bertacca:2017vod} independently of the bin width. This happens because the averaging procedure weights quantities w.r.t. the position of the sources, thus it does not affect very much the lensing contribution, which is generated by all the lenses encountered by the GWs along their path.
The RSD term increases with the bin width at high redshifts (i.e.,~$z\gtrsim1.2$), while it decreases at lower redshift. This happens because of the different behaviours of the contributions to the velocity term under the effect of the averaging procedure.
The velocity is indeed the sum of three contributions related to the peculiar velocity: the Kaiser term, proportional to $\ud v_\parallel/\ud \bar{\chi}$ in Eq.~\eqref{eq:total formula pois}, and two Doppler terms, usually referred as D1 and D2. The D1 term is proportional to $v_\parallel$, and the D2 term, which instead is proportional to $v$. The source functions of these three contributions have been given in Eq.~\eqref{eq:source_functions_Weighted_DL}. The Kaiser and D2 contributions grow with the size of the bin, whereas D1 diminishes. This different behaviour depends on both the different derivatives of the Bessel function inside the three terms and also the fact that the D1 term is the only one also present in the $\Delta\ln\D_L$ term (i.e. it enters twice in the definition of $\Delta\ln\DL$).
The D2 term is always subdominant, while a different combination of the Kaiser and D1 contributions at different redshifts and for different sizes of the bin determine the trend of the velocity spectrum discussed before. We show in appendix \ref{app:vel} a plot of the three contribution and how they change in the two bin width considered.
The anisotropies due to the perturbations of the potentials (i.e. the GR term in the $\texttt{CLASSgal}$ notation) are always subdominant.

\begin{figure}[h!]    
    \centering
    \includegraphics[width=\linewidth]{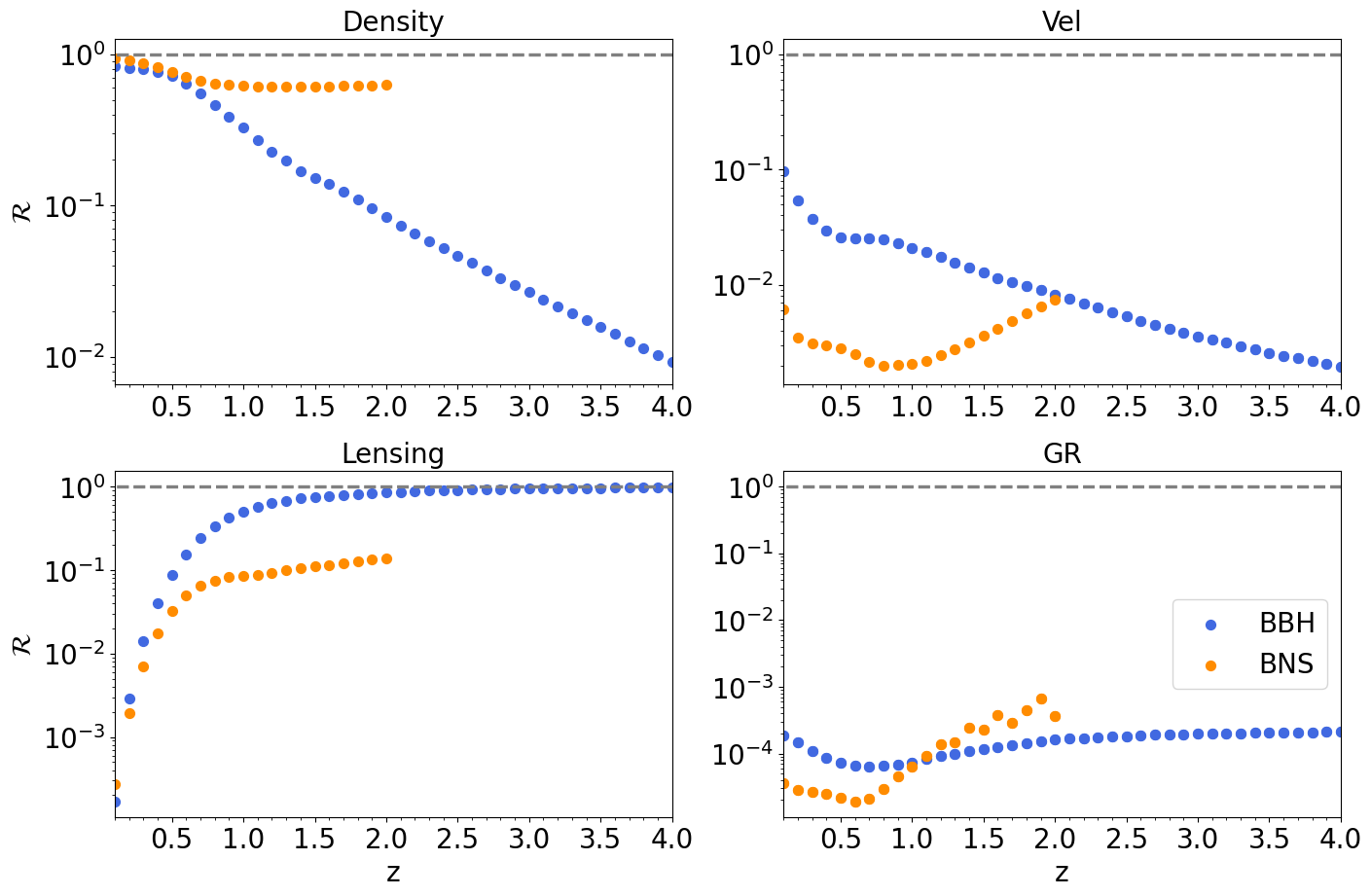}
    \caption{Plot of the ratios $\mathcal{R}$ of the density (upper left), velocity (upper right), lensing (lower left) and GR (lower right) corrections w.r.t. the total anisotropies, according to Eq.\eqref{eq:ratio} of the LD anisotropies $\Delta\ln\DL$ for ET+2CE. The density contribution is dominant at low redshifts while the lensing is dominant at higher redshifts. On the other hand, vel and GR are always subdominant.}
    \label{fig:ratio_ET2CE}
\end{figure}

\begin{figure}[t!]    
    \includegraphics[width=\linewidth]{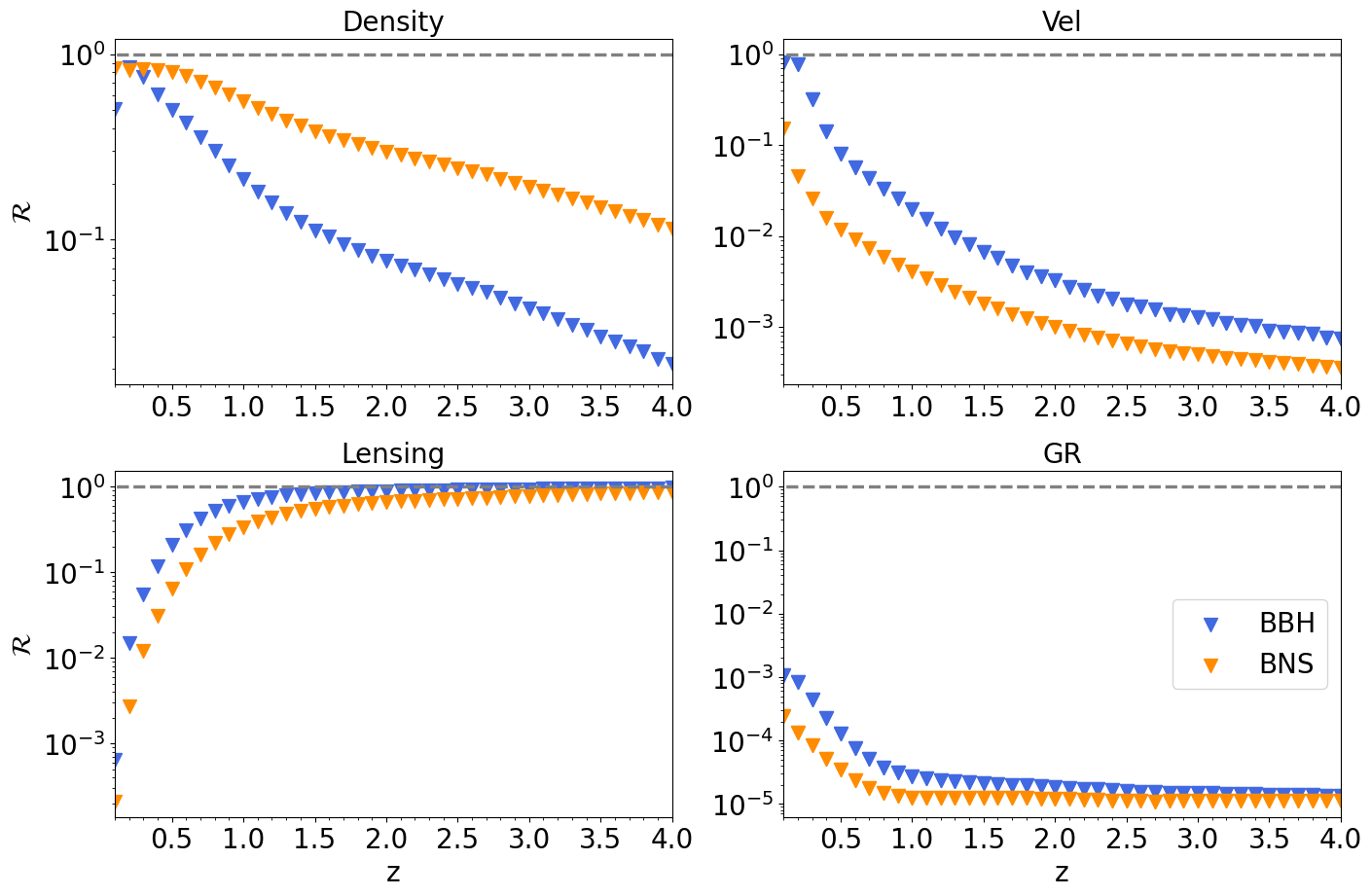}
    \caption{Plot of the ratios $\mathcal{R}$ of the density (upper left), velocity (upper right), lensing (lower left) and GR (lower right) corrections w.r.t. the total anisotropies, according to Eq.\eqref{eq:ratio} of the LD anisotropies $\Delta\ln\DL$ for BBO/DECIGO. We obtain a similar behaviour as the ET+CE case, so density dominates at low redshift, while at larger redshifts the lensing contribution is larger. The main difference is the larger relevance of the vel contribution due to the smaller bin width.}
    \label{fig:ratio_BBO}
\end{figure}

We now would like to see which of the different contributions (density, vel, lensing, GR) dominates and at which redshift, when summed over all the multipoles. We consider the ratio $\mathcal{R}_\alpha$ between the different contributions and the total intrinsic anisotropies,
\begin{equation}\label{eq:ratio}    \mathcal{R_\alpha}\equiv\frac{\sum_\ell(2\ell+1)C_{\ell\,\alpha}}{\sum_\ell(2\ell+1)C_{\ell}}\,,
\end{equation}
where $C_\ell$ is the total contribution (considering also the cross terms).
The binning is the one detailed in Appendix \ref{sec:Binning} and consists of equally spaced bins in $z$ while their width is determined by experimental uncertainties. Thus at each point of the plot we consider a bin width corresponding to $\sigma_{\D_L}$, which is the experimental precision for ET+2CE while it is 5 times the experimental precision for BBO/DECIGO. 
This choice was made to decrease the shot noise in the BBO/DECIGO case so to increase the SNR as we are going to see in Section \ref{Results}. We plot the $\mathcal{R}_\alpha$ in Figures \ref{fig:ratio_ET2CE}, \ref{fig:ratio_BBO} for ET+2CE and BBO/DECIGO respectively.

\subsection{The Kinematic Dipole }

\begin{figure}
    \centering
    \includegraphics[width=\linewidth]{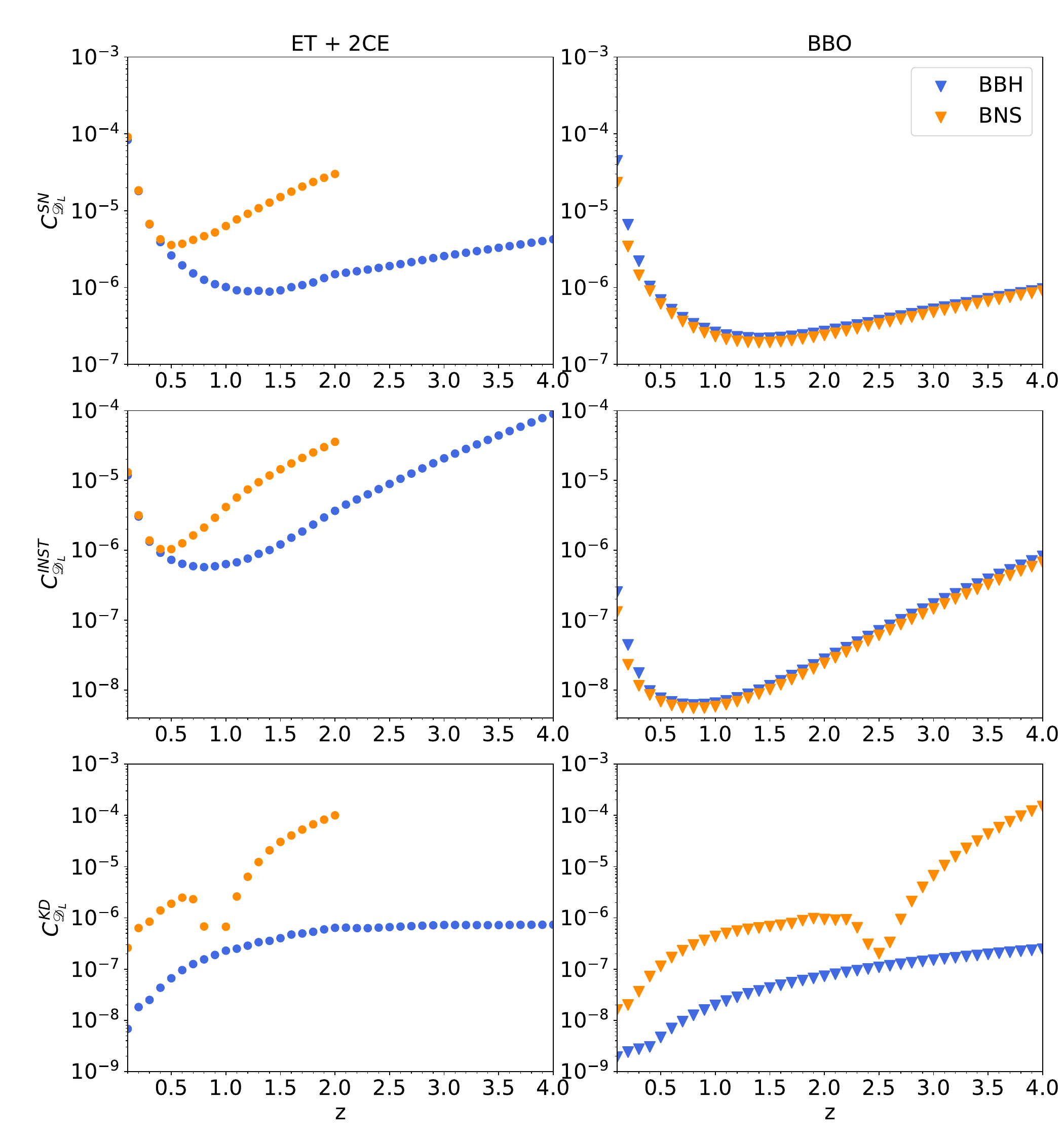}
    \caption{Plot of the amplitude of the shot noise (upper plots), instrumental noise (middle plots) and kinetic dipole (lower plots) contributions to $\Delta\ln\DL$ as a function of the center of the bin redshift for BBH (blue) and BNS (orange). In the plots on the left we have considered the binning of the experimental error of ET+2CE $\sigma_{\D_L}$, while in the plots on the right, we have considered BBO/DECIGO sensitivity while the sigma  of the bin is 5$\sigma_{\D_L}$, since this choice maximizes the SNR, as we will present later. 
    \label{fig:sn_instrn}}
\end{figure}

In Eq.~\eqref{eq:total formula pois}, several terms evaluated at the observer contribute to the anisotropies of $\DL$. It can be shown that the monopole contribution, proportional to 
$\Phi_o$, cannot be observed~\cite{Challinor:2011bk}. Therefore, among the observer terms, we consider only those proportional to the peculiar velocity of the observer, i.e.,~$v_{\|,o}$, which gives rise to a kinetic dipole (KD). The detectability of such a dipole for the number count of GW sources has been forecasted in~\cite{Mastrogiovanni:2022nya}, while for the astrophysical GW background in~\cite{ValbusaDallArmi:2022htu}. The source function of the KD, derived in Appendix~\ref{app:derivation of the kinetic dipole}, reads
\begin{equation} \begin{split}
    (\Delta\ln\DL)_{o}&=\vec{v}_{o}\cdot \hat{\n}^{\rm b}\int_{\overbar{V}}\ud\overbar{\x}^3\left[\WD\frac{1}{\overbar{\chi}\cH}+\left(\WD- \WV\right) \left(-\frac{5s-2}{\overbar{\chi}\cH}+2-b_{\rm e}+\frac{\cH'}{\cH^2}\right)\right]\, ,
\end{split} \end{equation}
where the peculiar velocity of the observer considered in this work is equal to the relative velocity of the Sun w.r.t. the CMB measured by \emph{Planck}~\cite{Planck:2013kqc}, $v_{\|,o}=(384 \pm 78 \pm 115) \, \rm km/s$. The angular power spectrum of the kinetic dipole, computed in Appendix~\ref{app:derivation of the kinetic dipole}, is equal to
\begin{equation} \begin{split}
    C_{\DL}^{\rm KD}&=\frac{4\pi}{9}\,v_o^2\, \delta_{\ell\,1}\left[\int\ud\overbar{\x}^3 \, \WD\frac{1}{\overbar{\chi}\cH}+\left(\WD-\WV\right)\left(-\frac{(5s-2)}{\overbar{\chi}\cH}+2-b_{\rm e}+\frac{\cH'}{\cH^2}\right)\right]^2 \, .
\end{split} \end{equation}
In Figure~\ref{fig:sn_instrn}, we have plotted the KD as a function of the center of the bin. As expected, in most of the redshift bins, the amplitude of the KD is larger than the intrinsic anisotropies, because of the large peculiar motion of the observer w.r.t. the LSS rest frame. In the case of the anisotropies of BNS detected with ET+2CE, the difference between $\WD$ and $\WV$ determines a compensation between different contributions to the KD which results in the minimum around $z=0.9$. In the case of BBO, the choice of a larger bin for BNS enhances the amplitude of the KD w.r.t. the ones of BBH.

\subsection{Shot noise and instrumental noise}
\label{sec:shot_noise}
\label{sec:instrumental_noise}
The estimator of the average LD, Eq.~\eqref{eq:def_DL_average_bin_W_c}, depends on the number of detected GW events. Since this discrete number is distributed according to a Poisson distribution, $\DL$ is affected by shot noise 
\begin{equation}
    n_{\rm GW}(\mathcal{D}_L,\hat{\n}) = \overbar{n}_{\rm GW}(\mathcal{D}_L)\left[1+\delta_{\rm SN}(\mathcal{D}_L,\hat{\n})\right] \, ,
    \label{eq:exp_N_GW_delta_SN}
\end{equation}
where the fluctuation of the density contrast $\delta_{SN}$ is a Gaussian random variable with mean and covariance given by 
\begin{equation}
    \begin{split}
   \left\langle  \delta_{\rm SN}(\mathcal{D}_L,\hat{\n})\right \rangle &=0  \, ,  \\
    \left\langle \prod_{i=1}^2 \delta_{\rm SN}(\mathcal{D}_{L,i},\hat{\n}_i)\,\overbar{n}_{\rm GW}(\mathcal{D}_{L,i})\,\ud V_{c,i}\right \rangle &=
    \overbar{n}_{\rm GW}\ud V_{c,1} \ud V_{c,2}\,  \delta(\D_{L,1}-\D_{L,2})\delta(\hat{\n}_1-\hat{\n}_2)   \, .    
    \label{eq:delta_sn_correlators}
    \end{split}
\end{equation}
In Appendix~\ref{app: shot noise}, we have shown that it is possible to write
\begin{equation}
    \begin{split}
        \DL= &  \BDL \left(1 + \int \ud \overbar{\x}^3 \left(\WD -\WV \right) \delta_{SN}\right) \, , 
    \end{split}
\end{equation}
therefore the perturbation of the shot noise $\sigma_{\rm SN}$ defined in Eq.~\eqref{eq:def_bar_DL_discrete} is just
\begin{equation}
    \ln \sigma_{\rm SN} = \int \ud \overbar{\x}^3 \left(\WD -\WV \right) \delta_{SN} \, .
\end{equation}
When two bins do not overlap, the shot noise is zero, because fluctuations in different volumes are uncorrelated. In this work we assume that the overlap of two window functions $W_{\rm b}$ with different $\hat{\n}^{\rm b}$ is always zero, since we have chosen the beam of Eq.~\eqref{def:cut_ell_max_instr} (or, equivalently, the maximum multipole in the analysis) in such a way that the separation of two bins is always larger than $\sigma_{\hat{\n}}$. This makes the overlap subdominant for the scope of this analysis. 
On the other hand, when the center of two bins are separated by a distance comparable with $\sigma_{\D_L}$, the shot noise of the correlation of two different shells could be nonvanishing, see, e.g., Eq.~\eqref{eq:sn_different_bins}. In the same shell the shot noise can be written as 
\begin{equation}
\begin{split}
    C^{\rm SN}_{\DL}&=\int \ud \overbar{\x}^3  \frac{1}{\overbar{n}_{\rm GW}}\left(\WD-\WV\right)^2 \, ,
    \label{eq:cov_sn}
\end{split}
\end{equation}
while we refer to Appendix~\ref{app: shot noise} for the full derivation and the cross bin formulation. In Figure~\ref{fig:sn_instrn} we have plotted the shot noise for BBH and BNS in the case of ET+2CE and BBO/DECIGO. The binning is the same as the one used previously. We take care of potential overlaps by computing the cross-covariance as detailed in Section \ref{Results}. At low redshift, the shot noise of BNS is smaller than the ones of BBH, because of the larger number of detected BNS in the local Universe. At larger redshift, the efficiency in the detection of BBH is much better, therefore less BNS are observed at $z\gtrsim 0.5$, thus the shot noise of BNS is much larger. In the case of BBO/DECIGO, we assume that all BBH and BNS could be resolved. Therefore the shot noise of BNS is always smaller than the one of BBH due to the larger number of BNS in the Universe.

One possibly counterintuitive characteristic of the shot noise of $\DL$ is that if the bin size increases the shot noise increases. This augment is driven by the amplification in $(\WD-\WV)^2$ in Eq.\eqref{eq:cov_sn}, which compensates the expected reduction due to a larger number of sources detected. We show this in Appendix \ref{app: shot noise}, where we plot in Figure \ref{fig:windows} the term $(\WD-\WV)^2$ considering two bin width. One can see that the larger bin size leads to a larger area under the curve.
One should also note that this reasoning works if the number density stays relatively constant inside the bins, so if there are many more source in the larger bin this would compensate the increase in the numerator of the shot noise.
The smaller sizes of the bins is indeed the main cause of improvement we obtain in studying $\DL $ with BBO/DECIGO, since this allows to reduce the shot noise, as can be seen from Figure \ref{fig:sn_instrn}.

We then evaluate the instrumental noise. The estimator of the average LD in a bin has been defined in Eq.~\eqref{eq:def_DL_average_bin_W_c}, weighting the data with a window function $W_{\rm b}$ which depends on the estimated LDs and angular positions of the individual sources. Since the single $\D_{L(i)}$, $\hat{\n}_{(i)}$ are affected by uncertainties due to instrumental noise (i.e. the experimental error represented by $\sigma_{\D_L}$ and $\sigma_{\hat{\n}}$ obtained from the Fisher analysis) 
, it is necessary to propagate these errors on $\DL$. 
The errors on the LD and the angular resolution of the individual sources have been plotted in Figure~\ref{fig:sn_instrn}, while details about their computation can be found in Appendix~\ref{app:population_source_and_instrumental_noise}.
Then the covariance of the autocorrelation in the continuous case is 
\begin{equation}
    \begin{split}
    C^{\rm instr}_{\DL} = \frac{1}{\left(\overbar{N}_{\rm GW}^{\rm b}\BDL\right)^2} \int \ud\overbar{\x}^3\, n_{\rm GW} \sigma_{\mathcal{D}_L}^2 \left\{\partial_{\mathcal{D}_L} \left[
     \left(\mathcal{D}_L-\BDL \right)W_{\rm b}
    \,\right]\right\}^2 \, , 
    \label{eq:cov_instr}
    \end{split}
\end{equation}
where $\partial_{\D_L}$ is the partial derivative w.r.t. the LD. This expression is consistent with the results obtained in~\cite{Namikawa:2015prh}. The multi-bin case and a step by step derivation can be found in Appendix \ref{app:instrumental_error}. The error on $\DL$ due to instrumental noise depends in principle also on $\sigma_{\hat{n}}$, but we have taken it into account in the beam function defined in Eq.~\eqref{def:cut_ell_max_instr}, which will determine a maximum angular scale at which we can measure the fluctuations $\Delta\ln\DL$, see e.g. Eqs.~\eqref{eq:SNR_pseudo_Cl},~\eqref{eq:SNR_pseudo_Cl_total}. In Figure~\ref{fig:sn_instrn} we plot the covariance of instrumental noise in the same bin for ET+2CE (middle left) and BBO/DECIGO (middle right). 
Considerations on the features of the spectra as a function of the redshift are analogous to the shot noise. While instead for the instrumental noise we recover the fact that it decrease when the bin size increases, in contrast with what we have found for the shot noise.

\section{Results}
\label{Results}

In this section, we discuss the detectability of the GR corrections on $\DL$ and of the kinetic dipole in presence of shot noise and instrumental noise. We will show results in the case of the detector network ET+2CE and for the more futuristic experiments BBO/DECIGO, by computing the signal-to-noise ratio (SNR).
The best estimator of the angular power spectrum in a single-bin are the ‘‘pseudo-Cls'', defined as
\begin{equation}
    \hat{C}_\ell \equiv \frac{1}{2\ell+1}\sum_m \left| \frac{\mathscr{D}_{L,\ell m}-\BDL}{\BDL}\right|^2 \, .
\end{equation}
The error on the pseudo-$C_\ell s$ is given by the combination of shot noise, instrumental noise and cosmic variance. 
We can then define the SNR as
\begin{equation}
    {\rm SNR}  \equiv  \left[\sum_{\ell}\frac{(2\ell+1) B_\ell^2C_{\ell}^2}{2\left(C_\ell+C_{\DL}^{\rm SN}+C_{\DL}^{\rm instr}+C_{\DL}^{\rm KD}\right)^2}\right]^{1/2} \, ,
    \label{eq:SNR_pseudo_Cl_total}
\end{equation}
where we included all corrections mentioned above (including the KD contribution) in Fig. \ref{fig:total_contrib}. In Appendix \ref{SNR_components} we also deal with the SNRs of the single components. In this Figure we can see that we could obtain a clear detection only in the BBO/DECIGO case, while we reach in the ET+2CE case an $\rm SNR\approx 1$ for BBH and $\rm SNR \approx 2$ at low redshifts for BNS.
As expected from Figure~\ref{fig:ratio_ET2CE}-\ref{fig:ratio_BBO}, the dominant contribution to the SNR is given by the lensing and the density terms at high and low redshifts respectively. For both ET+2CE and BBO/DECIGO, the shot noise is a bottleneck (in particular in the BBH case) in the detection of the angular power spectrum of GR effects, because of the low numbers of sources in the Universe.

Concerning ET+2CE we see that both BNS and BBH have $\rm SNR>1$ and the peak in the BNS case is mainly driven by the clustering contribution, while for BBH it is due to the lensing contribution.
In the BBO/DECIGO case, instead, we can see the characteristic shape due to the lensing contribution, while the difference between BNS and BBH at small redshifts is driven by the clustering contribution (which is larger for BNS).

Moreover, it is also interesting to quantify the entire amount of information that can be obtained by correlating the anisotropies of different bins. To do this, we consider all of the angular power spectra,
\begin{equation}
    \boldsymbol{\hat{C}_\ell} \equiv \begin{pmatrix}
        \hat{C}_\ell^{11} \\
        \hat{C}_\ell^{12} \\
        \vdots\\
        \hat{C}_\ell^{1N_{\rm bin}}\\
        \hat{C}_\ell^{22}\\ 
        \hat{C}_\ell^{23}\\
        \vdots\\
        \hat{C}_\ell^{N_{\rm bin}N_{\rm bin}}
    \end{pmatrix} \, ,
\end{equation}
with $N_{\rm bin}$ the total number of bins used. We have also introduced the compact notation
\begin{equation}
    \hat{C}_\ell^{12} \equiv \hat{C}_\ell\left(\D_{L,1}^{\rm b},\D_{L,2}^{\rm b}\right) \, . 
\end{equation}
To not count twice the information coming from the same couple of bins, the vector of observables $\boldsymbol{\hat{C}_\ell}$ has dimensions $N_{\rm bin}(N_{\rm bin}+1)/2$. Following~\cite{scelfo2018gw}, the covariance of the pseudo-Cls for different bins can be written as
\begin{equation}
    \begin{split}
    \Sigma^{12;34}_\ell \equiv&  {\rm cov}\left[\hat{C}_\ell\left(\D_{L,1}^{\rm b},\D_{L,2}^{\rm b}\right),\hat{C}_\ell\left(\D_{L,3}^{\rm b},\D_{L,4}^{\rm b}\right)\right]  =\frac{1}{2\ell+1}\left(C^{{\rm tot}\, , 12}_\ell C^{\rm tot\, , 34}_\ell+C^{{\rm tot}\, , 13}_\ell C^{\rm tot\, , 24}_\ell\right) \, ,
    \end{split}
\end{equation}
where we have defined
\begin{equation}
    C_\ell^{\rm tot\, , ij}\equiv C_\ell^{\rm ij}+C_{\DL}^{\rm SN\, , ij}+C_{\DL}^{\rm instr\, , ij}+C_{\DL}^{\rm KD\, , ij} \, . 
\end{equation}
The total SNR is defined by
\begin{equation}
    {\rm SNR}_{\rm global} \equiv \left[\sum_\ell \boldsymbol{\hat{C}_\ell}^T \boldsymbol{\Sigma}^{-1}_\ell \boldsymbol{\hat{C}_\ell}\right]^{1/2} = \left[\sum_\ell \sum_{i,j\geq i}\sum_{k,l\geq k}C_\ell^{ij}\left(\Sigma^{-1}\right)^{ ij;kl}_\ell C_\ell^{kl}\right]^{1/2} \, . 
\end{equation}
The choice of the widths of the bins for ET+2CE and BBO/DECIGO has been described in Appendix~\ref{sec:Binning}. The overall results are
\begin{equation}
    \begin{split}
        {\rm SNR}_{\rm global}^{\rm BBH\, , ET+2CE} = &\, 4.4 \, , \hspace{1.5em} {\rm SNR}_{\rm global}^{\rm BNS\, , ET+2CE} = \,6.6\, , \\
        {\rm SNR}_{\rm global}^{\rm BBH\, , BBO/DEC} = & 53 \, , \hspace{1.5em} {\rm SNR}_{\rm global}^{\rm BNS\, , BBO/DEC} = 79 \, .
    \end{split}
\end{equation}

As expected from the computation of the SNR in the single bins, the SNR in the case of BBO/DECIGO increases by more than one order of magnitude w.r.t. ET+2CE, because of the better angular resolution of the instrument, which allows to measure the angular power spectrum up to the non linear scale $\ell_{\rm nl}$ in most bins. The SNR for BNS is always larger than the one of BBH, because of the smaller shot noise due to the larger number of sources. The dominant contribution to the SNR is provided by lensing anisotropies, however also density perturbations and velocity effects could be detectable by correlating all the redshift bins. This is consistent with the results of the SNR in the individual bins plotted in Appendix \ref{SNR_components} in Figures~\ref{fig_SNR_map_ET2CE},~\ref{fig:SNR_Cls_BBO}. The SNR for BBH and BNS detected with ET+2CE and BBO/DECIGO in the multi-bin analysis are larger than two, therefore potentially providing additional and complementary information to e.g.,~ galaxy number counts or resolved GW events.

\begin{figure}[t!]
    \centering
\includegraphics[width=\linewidth]{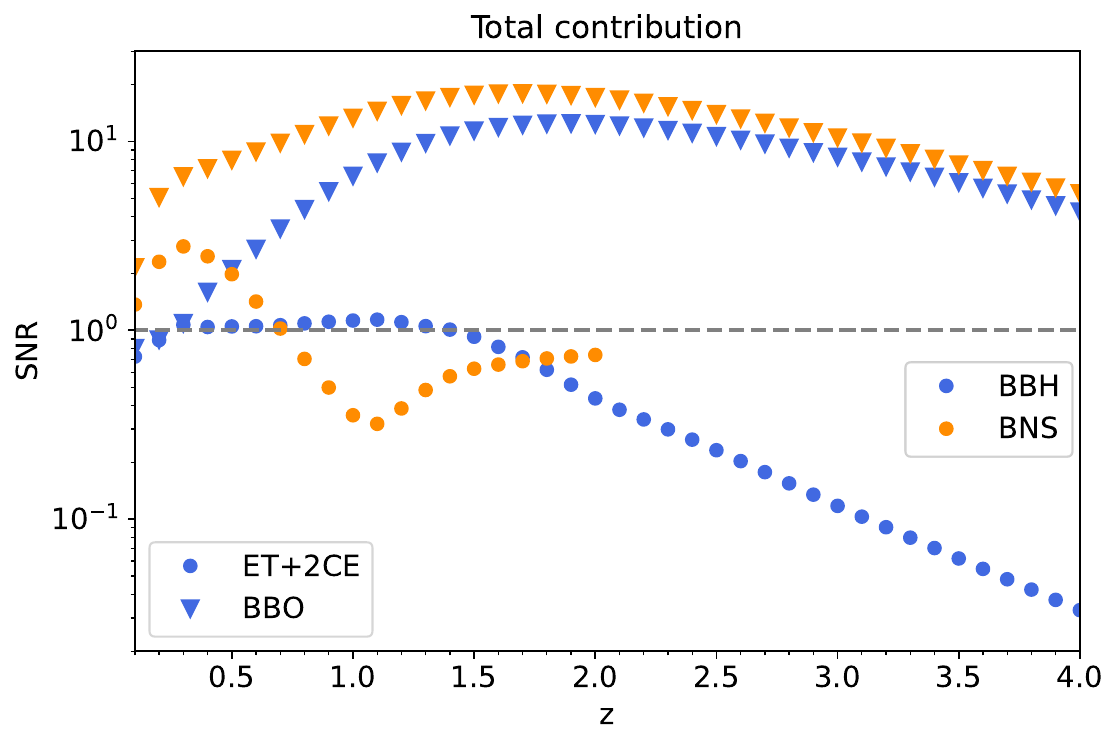}
    \caption{Plot of the SNR of the overall relativistic corrections and the KD from Eq.\eqref{eq:SNR_pseudo_Cl_total}. We see that we reach a SNR$>1$ for both BBH (blue) and BNS (orange) for ET+2CE (dotted). In both cases the signal at high redshifts is due to the KD. Regarding BBO (triangles), we see that for both types of sources we have a clear signal, mainly driven by the lensing contribution.}
    \label{fig:total_contrib}
\end{figure}

We now focus on the evaluation of the total error associated to GR corrections. This way GR corrections can be taken into account in the error budget of different analysis on the GW LD.
Following~\cite{Bertacca:2017vod}, we define the total error induced by the anisotropies as
\begin{equation}\label{sigma-anis}
    \frac{\sigma_{\rm anis}}{\D_L}\equiv\sqrt{\sum_\ell\frac{2\ell+1}{4\pi}B_\ell C_\ell}\, .
\end{equation}
We decided to not include the KD in this estimate since it can be obtain from Fig~\ref{fig:sn_instrn}.
\begin{figure}
    \includegraphics[width=\linewidth]{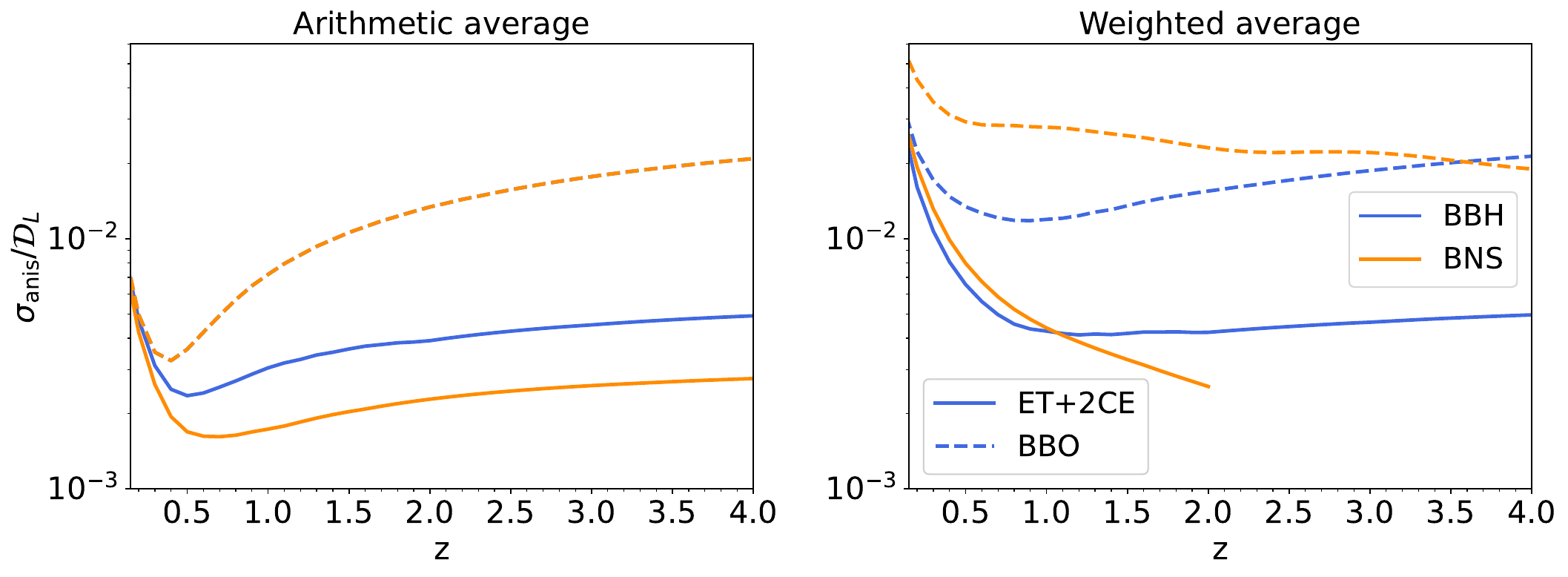}
    \caption{Plot of the contribution to $\ln\DL$ for the arithmetic average (left) and weighted average (right) due to GR corrections as a function of the bin's redshift.}
    \label{fig:fitting_plot}
\end{figure}
We also provide in appendix \ref{app:fitting_formula} the fits to reobtain the curves of Figure \ref{fig:fitting_plot}.

\section{Conclusions}
\label{Conclusions}
In the future, measurements of the luminosity distance of gravitational wave sources could be a powerful tool to probe GR and cosmology. While the effects of cosmological perturbations and projection effects on the luminosity distance (LD) of the single sources have been evaluated before, in this work we compute, for the first time, those effects for the case of the average LD of many ($10^4-10^5$ per year) detected sources. Future detectors will observe a large number of GW, thus it becomes important to account for their clustering and for averaging. The weighted LD is obtained by convolving the LD of the single sources with a window function that weights the GW events inside the volume. 
In this paper we focus on the computation intrinsic (density, kinematic and GR) anisotropies which are generated by the fluctuations of the LD of the individual events computed in~\cite{Bertacca:2017vod} and by perturbations in the number of mergers in the volume. The kinetic dipole is produced by our peculiar motion w.r.t. the rest frame of the sources, while the shot noise is due to Poisson fluctuations of the discrete number of events in the bin. We consider the shot and instrumental noise anisotropies as foregrounds for the detection of the intrinsic terms.

The main result of this work is the analytical expression of the GR corrections on the weighted LD, computed in a gauge-free setting considering a perturbed FLRW Universe. The GR corrections have been computed in Eq.~\eqref{eq:total formula pois} in the Poisson gauge and in Eq.~\eqref{GR_correction_Delta_DL_generic_gauge} in a generic gauge. The angular power spectrum of the GR corrections have been calculated by using a modified version of \texttt{Multi$\_$CLASS}, finding results in agreement with~\cite{Bertacca:2017vod} in the limit of thin bins. As expected, the dominant contribution to the anisotropies is given by the lensing term at high redshifts, while the density perturbation dominates at low redshift and when the bin width is large in $z$ (these conditions are verified in the BNS case).  
Expressions for the covariance matrix of the corrections on the average LD due to shot noise and instrumental noise have been presented in Eqs.~\eqref{eq:cov_sn},~\eqref{eq:cov_instr} respectively. 

We have estimated the detectability of the angular power spectrum of the GR corrections and the KD 
on the average LD of BBH and BNS with future ground- and space-based interferometers ET+2CE and BBO/DECIGO. We have computed the SNR of the intrinsic anisotropies of the weighted LD by considering one bin and many bins at the same time in the analysis. In the ET+2CE case, the shot noise and the instrumental errors are too large to make GR contributions relevant on the average LD on the single bin, given the poor angular resolution and low number of merging BBH and BNS detected per year. In a future work we will explore the possibility of reducing the impact of shot noise errors by using a multi-tracer and map-based analysis.
In the case of BBO/DECIGO, our results show that it is possible to detect the density correction to the weighted LD of BNS with $\rm SNR>2$ in the large majority of the single bins considered, reaching the maximum signal to noise at $z\approx 1$, where it is close to 7. This can be achieved because of the high angular and LD resolutions of these interferometers and the large number of detected sources. The total SNR we get for ET+2CE is 6.6 and 4.4 for BNS and BBH respectively, while for BBO the detactability improves roughly by an order of magnitude, finding SNR of the order of 79 and 53.
The large SNR of the GR corrections estimated for BBO/DECIGO or for BBH in the case of ET+2CE implies measurements with high significance of the GR corrections of the weighted LD $\DL$
could provide a new powerful source of information that could be used in cross-correlation with LSS probes to obtain information about GR and cosmology.

\section*{Acknowledgments}
The authors thank Sabino Matarrese and Angelo Ricciardone for useful discussions. A.~B, L.~V. and A.~R thank Sarah Libanore for helpful comments. D.~B. acknowledges partial financial support from the COSMOS
network (www.cosmosnet.it) through the ASI
(Italian Space Agency) Grants 2016-24-H.0, 2016-24-H.1-2018 and
2020-9-HH.
AR acknowledges
funding from the Italian Ministry of University and Research (MUR) through the “Dipartimenti di eccellenza” project “Science of the Universe”.
L.~V. acknowledges financial support from the Supporting TAlent in Re- Search@University of Padova (STARS@UNIPD) for the project “Constraining Cosmology and Astrophysics with Gravitational Waves, Cosmic Microwave Background and Large-Scale Structure cross-correlations”. 
This work is partially supported by ICSC – Centro Nazionale di Ricerca in High Performance Computing, Big Data and Quantum Computing, funded by European Union – NextGenerationEU.

\appendix

\section{Population of the sources}
\label{app:population_source_and_instrumental_noise}

In this work we consider a population of BBH and BNS of stellar origin, which can be detected by the third-generation interferometers ET and CE and onwards (e.g. BBO and DECIGO). To characterize the distribution of the intrinsic parameters of the binaries, we use the most recent constraints provided by LVK~\cite{LIGOScientific:2021djp}. For the distribution of the masses of BBH we consider the Power Law + Peak mass model, between $2.5$ and $100\, M_\odot$, while for BNS we use a uniform distribution between $1$ and $2.5\, M_\odot$. The spin of BBH have been sampled according to a Gaussian distribution with zero average and error equal to $0.1$, while in the case of BNS taking the spins equal to zero seems to be a good approximation~\cite{Iacovelli:2022bbs}. In the case of BBH, we have considered zero tidal deformability, while for BNS we have considered a uniform distribution between $0$ and $2000$. We have also adopted a uniform distribution for the inclination angle of the binaries. The redshift distribution of BBH and BNS has been evaluated by using UniverseMachine~\cite{Behroozi_2019} as in~\cite{Bellomo:2021mer}. To compute the merger rate of the binaries we considered the average SFR per halo and we convolved it with the halo mass function of~\cite{Tinker:2008ff} and a time delay distribution between the formation of the binary and the merger. We used $p(t_d)$ with a minimum time delay of $50\, \rm Myr$ for BBH and $20 \, \rm Myr$ for BNS. The merger rate obtained have been normalized to the local number of detections of BBH and BNS, $R_{\rm local}^{\rm BBH}=28.3\, {\rm yr}^{-1}\, {\rm Gpc}^{-3}$ and $R_{\rm local}^{\rm BNS}=105.5\, {\rm yr}^{-1}\, {\rm Gpc}^{-3}$~\cite{LIGOScientific:2021djp}. To compute the waveforms, we used $\texttt{IMRPhenomHM}$ for BBH and $\texttt{IMRPhenomD\_NRTidalv2}$ for BNS.

Once the population of the sources have been fully characterized, we quantified the detectability of the individual events by the network ET+2CE and BBO/DECIGO. In the former case, we considered ET with triangular configuration and sensitivity ET-D~\cite{ETsens} located in Sardinia, while we considered one CE in the US and another one in Australia, with sensitivities given in~\cite{CEsens}. The SNR of a binary which emits GWs at redshift $z_i$, with intrinsic parameters $\vec{\theta}_i$ is given by
\begin{equation}
    \varrho_i = \left[4\int \ud f \sum_A\frac{|h_A(\hat{\n}_i,\vec{\theta}_i,z_i,f)|^2}{N_{AA}(f)}\right]^{1/2} \, ,
\end{equation}
where $N_{AA}(f)$ is the power spectral density (PSD) of the noise of the interferometer $A$ in the network, while $h_A(\hat{\n}_i,\vec{\theta}_i,z_i,f)\equiv \sum_\alpha F_A^\alpha(\hat{\n}_i,f)\, h_\alpha(\hat{\n}_i,\vec{\theta}_i,z_i,f)$, with $F_A^\alpha$ the detector pattern function for the polarization $\alpha$. In this work, we say that a source is resolved when $\varrho_i\geq \varrho_{\rm thres}$. In this work we set $\varrho_{\rm thres}=12$ and we select the events in the full population whose SNR is larger than the threshold. The results are depicted in Figure~\ref{fig:rho_threshold_quantities}. Once we have simulated the population of detected BBH and BNS, we evaluate the average error on the LD and angular resolution of the sources. To do this, we have computed the Fisher matrix for the GW events in the catalog, defined by
\begin{equation}
    \mathcal{F}^i_{\alpha\beta}\equiv \mathcal{F}_{\alpha\beta}(\hat{\n}_i,\vec{\theta}_i,z_i,f) = 4\int \ud f \sum_A\frac{\partial_\alpha h_A(\hat{\n}_i,\vec{\theta}_i,z_i,f)\partial_\beta h^*_A(\hat{\n}_i,\vec{\theta}_i,z_i,f)}{N_{AA}(f)} \, , 
\end{equation}
where the derivatives have been computed w.r.t. the parameters $\vec{\theta}$, $\D_L(z)$ and $\hat{\n}$. The Fisher matrix has been evaluated with $\texttt{GWFAST}$~\cite{Iacovelli:2022mbg}. To compute the covariance on the intrinsic parameters and on the position of the binaries, we invert the Fisher matrix,
\begin{equation}
    \mathcal{C}^i_{\alpha\beta}\equiv  \mathcal{F}^{-1\, i }_{\alpha\beta} \, .
\end{equation}
In this work, we are interested in studying the average LD of GW events, thus we disregard any information about the intrinsic parameters of the binaries $\vec{\theta}$, because two binary systems with different masses, spins or inclination angles are considered the same kind of object in this work. Therefore, we marginalize the uncertainties in the Fisher matrix of the parameters of the single binaries w.r.t. $\vec{\theta}$ and to compute the errors on the angular resolution and LD we consider just the elements in the inverse of the Fisher matrix which correspond to the derivatives w.r.t. $\D_L(z)$ and $\hat{\n}$,
\begin{equation}
    \mathcal{C}_{\alpha\beta,\rm marg}^i \equiv \begin{pmatrix}
    \mathcal{F}^{-1\, i}_{\D_{L} \D_{L}} & \,    \mathcal{F}^{-1\, i }_{\D_{L} \hat{\n}} \\[6pt]
    \mathcal{F}^{-1\, i }_{\D_{L} \hat{\n}} & \,    \mathcal{F}^{-1\, i}_{\hat{\n}\hat{\n}} \, .
    \end{pmatrix}
\end{equation}
From the marginalized covariance matrix of the individual events, we build an estimate of the average error at a given LD $\D_L$. To do this, we compute the average covariance between the LD and the angular resolution for all the events which has $\D_{L,i}$ in the range $I(\D_L)\equiv \left[\D_L-\Delta\D_L,\D_L+\Delta \D_L\right]$, with $\Delta\D_L\approx 0.1\, \D_L$. The average errors on the LD and angular resolution are defined then by 
\begin{equation}
    \begin{split}
        \sigma^2_{\D_L}(\D_L) \equiv& \frac{1}{N_{\D_L}}\sum_{i}  \mathcal{C}_{\D_L\D_L,\rm marg}^i \, , \\
        \sigma^2_{\hat{\n}}(\D_L) \equiv &  \frac{1}{N_{\D_L}}\sum_{i}  \mathcal{C}_{\hat{\n}\hat{\n},\rm marg}^i \, ,
    \end{split}
\end{equation}
where $i$ runs over the events with $D_{L,i}\in I(\D_L)$ and $N_{\D_L}$ is the number of objects with $D_{L,i}\in I(\D_L)$. In the continuous limit, they correspond to the following averages,
\begin{equation}
    \begin{split}
        \sigma^2_{\D_L}(\D_L) =& \int \ud \vec{\theta} \, p(\vec{\theta})\, \mathcal{C}_{\D_L\D_L}(\hat{\n},\vec{\theta},z,f) \; , \\
        \sigma^2_{\hat{\n}}(\D_L) =& \int \ud \vec{\theta} \, p(\vec{\theta}) \, \mathcal{C}_{\hat{\n}\hat{\n}}(\hat{\n},\vec{\theta},z,f) \; ,
    \end{split}
\end{equation}
with $p(\vec{\theta})$ the distribution of the intrinsic parameters of the binaries described at the beginning of this appendix. As stressed before, in this work we are not interested in distinguishing binary systems with different intrinsic parameters, thus, although BBH and BNS with different $\vec{\theta}$ have different errors on the LD and the angular position, we consider just the average errors of the full population. 
Moreover, in general the error on the LD and on the angular resolution are correlated, but we have found that this correlation is subdominant and not important for the purposes of our paper. The results for BBH and BNS for ET+2CE are shown in Figure~\ref{fig:rho_threshold_quantities}. 
As said in the main text of the total population the detected over the threshold (in this work we take $\rho_{\rm thres}=12$) consist to 93\% and 43\% respectively of the full BBH and BNS populations. As we highlighted in the main text we want to make a cut which is also linked to the precision of the measurements. We select from the data the events with good angular resolution, $\sigma_{\hat{\n}}<20 \, \rm deg^2$ for BBH and $\sigma_{\hat{\n}}<40 \, \rm deg^2$ for BNS, because they have on average a smaller SNR. After applying this cut we remain with 65\% and 11\% of the total population, respectively for BBH and BNS.  
Concerning the BBO/DECIGO case we assumed that all the sources are detected~\cite{Cutler:2005qq,Harms:2008xv} (up to the maximum redshift considered in this work which is $z=4$) and we considered the error on the LD and angular resolution computed in~\cite{Cutler:2009qv}. 
The bias of the GWs, introduced in Eq.~\eqref{eq:total formula pois}, has been computed~\cite{Bellomo:2021mer} by tracing the GWs by looking at the average SFR per halo computed in~\cite{Behroozi_2019}, $\langle {\rm SFR}_{\rm SF}$, and by using the halo bias of~\cite{Tinker_2010}, 
\begin{equation}
    b_{\rm GW}(\D_L) = \frac{\int \ud M_h \frac{\ud \overbar{n}_{\rm GW}}{\ud M_h}(M_h,\D_L) b_h(M_h,\D_L)}{\int \ud M_h \frac{\ud n}{\ud M_h}(M_h,\D_L)}\, ,
\end{equation}
where the number of GWs in all the halos of mass $M_h$ has been defined by using
\begin{equation}
    \frac{\ud\overbar{n}_{\rm GW}}{\ud M_h}(M_h,\D_L)\equiv N^{\rm LVK}_{\rm GW}\frac{\int \ud t_d \, p(t_d)\, \frac{\ud n}{\ud M_h}\left(M_h,\D_L^d\right)\langle {\rm SFR}_{\rm SF}\rangle\left(M_h,\D_L^d\right)}{\int \ud t_d \, p(t_d)\, \frac{\ud n}{\ud M_h}\left(M_h,0\right)\langle {\rm SFR}_{\rm SF}\rangle\left(M_h,0\right)}\, ,
\end{equation}
with $\D_L^d$ the LD of the binary at the formation and $t_d$ the time delay between the formation of the binary and the merger, distributed with $p(t_d)\sim 1/t_d$ from $50\, \rm Myr$ and the age of the Universe.

\section{Definition of useful quantities in Cosmic Ruler formalism}
\label{app:Cosmic Rulers}

In Section~\ref{The Weighted Luminosity Distance} we computed the GR corrections to the weighted LD within the ‘‘Cosmic Ruler framework''~\cite{Jeong:2011as,Schmidt:2012ne}. In this appendix we give more context to the formalism we adopted, while in the next one we compute the explicit expressions of the source terms of the perturbations.
We first introduce the metric $ \ti g_{\mu\nu}$, over which the GWs propagate, that takes the general form of
\begin{equation}\begin{split}\label{metric}
\ud s^2=a^2(\eta)[-(1+2A)\ud\eta^2-2B_i\ud\eta\ud x^i+(\delta_{ij}+C_{ij})\ud x^i\ud x^j]=\ti g_{\mu\nu}\ud x^{\mu}\ud x^{\nu}=a^2\hat{g}_{\mu\nu}\ud x^{\mu}\ud x^{\nu}\,,
\end{split}\end{equation}
where $a(\eta)$ is the scale factor as function of the comoving time $\eta$ and we introduced the comoving metric $\hat{g}_{\mu\nu}$. Moreover we also have that $A$ is a scalar, $B_i = \p_i B + \hat{B}_i$, where $B$ is a scalar and $\hat{B}_i$ is a solenoidal vector, i.e. divergence-free. Then $C_{ij}= 2D\delta_{ij} + F_{ij}$ where $F_{ij}=(\p_i\p_j-\delta_{ij}\nabla^2/3)F + \p_i\hat{F}_j + \p_j\hat{F}_i + \hat{h}_{ij}$, where $D$ and $F$ are scalars, $\hat{F}_i$ is a solenoidal vector and $\p^i\hat{h}_{ij}=\hat{h}^i_i=0$ a tensor perturbation. From now on we will use only the comoving metric $\hat{g}_{\mu\nu}=\ti g_{\mu\nu}/a^2$, so all the geodesic equations will be solved in this metric and also the affine parameter will be defined accordingly. 
In the CL frame the comoving distance and the unit vector associated to the direction of observation, introduced in Eq.\eqref{eq:coo_of_events}, are
\begin{equation}
    \begin{split}
    \overbar{\chi}=\sqrt{\overbar{x}^i\overbar{x}_i}\, , \hspace{2em} n^i=\frac{\overbar{x}^i}{\overbar{\chi}}=\delta^{ij}\frac{\p\overbar{\chi}}{\p\overbar{x}^j}\, .
    \end{split}
\end{equation}
In this analysis, $\overbar{\chi}$ is used as an affine parameter in the CL frame to describe the past GW-cone. 
This means that the total derivative w.r.t. $\overbar{\chi}$ is
\begin{equation}\begin{split}\label{total der}
\frac{\ud}{\ud\overbar{\chi}}=-\frac{\p}{\p\overbar{\eta}}+n^i\frac{\p}{\p\overbar{x}^i}
\end{split}\end{equation}
Moreover we define the directional derivatives as
\begin{equation}\begin{split}
\bb\p\pp=n^i\frac{\p}{\p\overbar{x}^i}\,,\qquad \bb\p\pp^2=\bb\p\pp\bb\p\pp\,,\qquad \bb\p_{\perp i}=\Per^j_i\bb\p_j\,,\qquad
\frac{\p n^j}{\p \overbar{x}^i}=\frac{\Per^j_i}{\overbar{\chi}}\,.
\end{split}\end{equation}
 Now we need define a map 
from the real to the CL frame in which observations are performed.
For a given GW path, we define this map as
\begin{equation}\begin{split}
x^\mu(\chi(\overbar \chi))=\overbar{x}^{\mu}(\overbar \chi)+\Delta x^{\mu} (\overbar \chi)
\end{split}\end{equation}
where
\begin{equation}\begin{split} \label{Deltax}
\Delta x^{\mu} (\overbar \chi) =  \frac{\ud  \overbar{x}^\mu }{\ud \overbar \chi} \delta \chi+ \delta x^{\mu} (\overbar \chi) \quad {\rm and} \quad \chi = \overbar \chi+ \delta \chi \;.
\end{split}\end{equation}
The GW 
is described by the null geodesic vector $k^\mu = \ud x^\mu / \ud \chi$ and in the  CL frame it is defined as
\begin{equation}\begin{split}
\label{kmu-0}
 \overbar{k}^\mu=\frac{\ud  \overbar{x}^\mu }{\ud \overbar \chi}=\left(-1, \; n^i \right)\;,
\end{split}\end{equation}
while, in the physical frame, $k^\mu$ evaluated at $\overbar \chi$ is 
\begin{equation}\begin{split}
\label{kmu}
 k^\mu(\overbar \chi) = \frac{\ud  x^\mu }{\ud \overbar \chi}(\overbar \chi)= \frac{\ud }{\ud \overbar \chi}  \left(\overbar{x}^\mu + \delta x^\mu\right)(\overbar \chi) = \left(-1+\delta \nu, n^i+\delta n^{i} \right)(\overbar \chi)=\overbar{k}^{\mu}+\delta k^{\mu}\,,.
\end{split}\end{equation}
where we are allowed to write $k^{\mu}$ as function of $\overbar{\chi}$ since it is equal to $k^{\mu}(\chi)$ at first order and $\delta x^{\mu}=x^{\mu}-\overbar{x}^{\mu}$, where both terms are evaluated at $\overbar{\chi}$, $\delta\nu$ is the perturbation of the zero-th component, while $\delta n^i$ is the perturbation of the spatial part. 
In the definition of $\delta x^{\mu}$ both the two coordinates are described with respect to $\overbar{\chi}$ (i.e. we are in the CL frame). On the other hand $\Delta x^{\mu}$ is a bridge quantity between the real space and the CL background and this way we can set a map, valid at linear order, between the two frames as
\begin{equation}\begin{split}\label{def of Delta}
\Delta x^{\mu}(\overbar{\chi})=\overbar{k}^{\mu}\delta \chi+\delta x^{\mu}\,.
\end{split}\end{equation}
From Eq.(\ref{kmu}) we obtain:
\begin{equation}\begin{split}\label{def delta x}
\delta x^0(\overbar{\chi})=\delta x^0_o+\int^{\overbar{\chi}}_0\ud\ti\chi\delta\nu(\ti\chi)\,,\qquad \delta x^i(\overbar{\chi})=\delta x^i_o+\int^{\overbar{\chi}}_0\ud\ti\chi\delta n^i(\ti\chi)\,,
\end{split}\end{equation}
where $\delta x^0$ and $\delta x^i$ are the shift of the time and spatial coordinates. In full generality we included the terms arising from the perturbation at the observer, indicated with the subscript  ``$o$".
These terms can be written as \cite{Fanizza:2018qux, Biern:2016kys, Bertacca:2019fnt}
\begin{equation} \begin{split}\label{eq:pert_at_the_observer}
\delta x^0_o=-\int^{\overbar{\eta}_0}_{\overbar{\eta}_{\rm in}} \overbar{a}(\tilde{\eta}) A(\tilde{\eta},\mathbf{0})\ud\tilde{\eta}\\
\delta x^i_o=\int^{\overbar{\eta}_0}_{\overbar{\eta}_{\rm in}}\ud x^i=\int^{\overbar{\eta}_0}_{\overbar{\eta}_{\rm in}}v^i\ud\tilde{\eta} \\
\delta a_o=\cH_0\delta x^0_o=-\cH_0\int^{\overbar{\eta}_0}_{\overbar{\eta}_{\rm in}} \overbar{a}(\tilde{\eta}) A(\tilde{\eta},\mathbf{0})\ud\tilde{\eta}\,,
\end{split} \end{equation}
where we also included the definition of the scale factor perturbations at the observer.
In the first definition one can see that, as expected, the perturbation of the 00 metric component (i.e. $A$) at the observer produces the time coordinates shift, while the source of the spatial coordinate shift is the velocity of the observer. Finally in the case of the scale factor, the perturbation at the observer is obtained from the time coordinates shift times the comoving Hubble constant. In the previous equations we integrated from an initial comoving time indicated with $\overbar{\eta}_{\rm in}$.
All calculations in this and the following section are performed in a gauge-free setting. We write the geodesic equations for the basic quantities $\delta\nu$ and $\delta n^i$. The observer is placed at $\overbar\chi=0$ and it is assumed to follow a geodesic motion comoving with the cosmic fluid. We choose them to lie at the spatial origin 0, and to observe at a proper time $\eta_0$.  
The geodesic equation is written as
\begin{equation}\begin{split}\label{geo eq k}
\frac{\ud k^{\mu}(\chi)}{\ud\chi}+\hat{\Gamma}^{\mu}_{\alpha\beta}( x ^{\gamma}) k^{\alpha}(\chi) k^{\beta}(\chi)=0\,,
\end{split}\end{equation}
where $\hat{\Gamma}^{\mu}_{\alpha\beta}( x ^{\gamma})$ are the Christoffel symbols referred to the comoving metric.
We can rewrite this equation, in a more manageable way for our aims, if we expand it at first order in $\overbar{\chi}$
\begin{equation}\begin{split}
\frac{\ud \delta k^{\mu}}{\ud\overbar{\chi}}+\delta\hat{\Gamma}^{\mu}_{\alpha\beta}(\overbar{x} ^{\gamma})\overbar{k}^{\alpha}(\overbar{\chi})\overbar{k}^{\beta}(\overbar{\chi})=0\,.
\end{split}\end{equation}
One need then to introduce the definition of GW
and from this one we can write all the terms inside the geodesic equation, a complete derivation can be found in \cite{Bertacca:2017vod}.

\section{Relativistic contributions terms}
\label{app:Relativistic contributions terms}

In this appendix we compute all the linear contributions to the GR corrections on $\DL$. To do this, we start from the definition of weighted LD given in Eq.\eqref{eq:def_DL_GR}. After some straightforward calculations, it is possible to re-write $\DL$ as the following first-order expansion in the map between the two frames,
\begin{equation} \begin{split}
     \mathscr{D}_L&=\frac{\int\ud\overbar{\x}^3\, W_{\rm b}\BD_L\overbar{n}^{\rm ph}_{\rm GW}\left(1+3\Delta\ln a+\frac{1}{2}\delta\hat{g}^{\mu}_{\mu}+\Delta\ln\D_L+\Delta\ln n_{\rm GW}^{\rm ph}+\Delta V\right)}{\int\ud\overbar{\x}^3\, W_{\rm b}\, \overbar{n}^{\rm ph}_{\rm GW}\left(1+3\Delta\ln a+\frac{1}{2}\delta\hat{g}^{\mu}_{\mu}+\Delta\ln n_{\rm GW}^{\rm ph}+\Delta V\right)}\\
    &=\overbar{\mathscr{D}}_L\left[1+\int\ud\overbar{\x}^3 \,\mathcal{W}_{\D_L}\Delta\ln\D_L+\left(\mathcal{W}_{\D_L}- \mathcal{W}_V\right)\,\Delta_{\rm GW}\right]\, .\label{calcolo 1 DL}
\end{split} \end{equation}
Here we will present the contributions in a gauge-free settings considering all the terms at the observer, while in the main text we presented only the final expression of $\DL$ in the Poisson gauge. The perturbation of the LD of a single source was computed in the ‘‘Cosmic Ruler'' framework in the Poisson gauge in~\cite{Bertacca:2017dzm}, while here we extended the derivation to the gauge-free setting
\begin{equation} \begin{split}
    \Delta\ln\D_L&=-\hat{\kappa}+\frac{1}{4}\Per^{ij}C_{ij}+\left( 1-\frac{1}{\overbar{\chi}\cH}\right)\Delta\ln a-\frac{1}{\overbar{\chi}}(T-\delta x_{\| o}-\delta x^0_o)\,.
\end{split} \end{equation}
The perturbation of the volume element at first order is (e.g. \cite{Scaccabarozzi:2018vux,Bertacca:2014wga})
\begin{equation} \begin{split}
    \Delta V&=v\pp-B\pp-\frac{1}{2}C\pp-\frac{2T}{\overbar{\chi}}-2\hat{\kappa}+\frac{2}{\overbar{\chi}}\left( \delta x_{\| o}+\delta x^0_o\right)-\left ( \frac{\cH'}{\cH^2}+\frac{2}{\overbar{\chi}\cH}+\frac{1}{\cH}\frac{\ud}{\ud\overbar{\chi}}\right)\Delta\ln a \,.
\end{split} \end{equation}
Moreover we introduce the perturbation to the metric and to the scale factor, and the latter will become useful in the perturbation to the physical number density of sources
\begin{equation} \begin{split}    
    \frac{1}{2}\delta\hat{g}^{\mu}_{\mu}&=A+\frac{1}{2}C^i_i \\
    \Delta \ln a &=v\pp+\delta a_o- v_{\parallel ,o} +A_o-A+2I\,,
\end{split} \end{equation}
The last quantity we need to introduce is the perturbation to the physical number density 
\begin{equation} 
    \begin{split}
    \Delta\ln n_{\rm GW}^{\rm ph}&=\delta_{\rm GW}-5s\left[-\hat{\kappa}+\frac{1}{4}\Per^{ij}C_{ij}+\left( 1-\frac{1}{\overbar{\chi}\cH}\right)\Delta\ln a-\frac{1}{\overbar{\chi}}(T-\delta x_{\| o}-\delta x^0_o)\right]\\
    &\,\,\,\,\,\,+\frac{\p \ln\overbar{n}_{\rm GW}}{\p\ln\overbar{a}}\Delta\ln a \,,
    \end{split} 
\end{equation}
which is affected by the usual density perturbation $\delta_{\rm GW}= n(\overbar{\x})/\overbar{n}(\overbar{\x})-1$, a term which is the result of the magnification of the sources (i.e. proportional to $s$) and a term proportional to $\Delta\ln a $ representing the evolution of the number of sources across redshift.
We also introduced in the gauge-free setting 
\begin{equation} \begin{split}
I=-\frac{1}{2}\int_0^{\overbar{\chi}}\ud\ti\chi\left (A' - B'\pp -\frac{1}{2}C'\pp  \right)\,,\quad T=-\int^{\overbar{\chi}}_0 \ud\ti\chi\left(A-B\pp -\frac{1}{2}C\pp\right)\,,
\end{split} \end{equation}
where $I$ is the Integrated Sachs Wolf effect~\cite{Sachs:1967er} and $T$ is the Shapiro time delay~\cite{Shapiro:1964uw}.
Moreover $\hat{\kappa}$ is the weak lensing convergence, defined as \cite{Jeong:2011as,Hirata:2010ba}
\begin{equation} 
    \begin{split}
        \hat{\kappa} &=-\frac{1}{2}\overbar{\p}_{\perp i}\Delta x_{\perp}^i=-\frac{1}{2}\overbar{\p}_{\perp i}\delta x^i_{\perp}\\
        &=\frac{\delta x_{\|,o}}{\overbar{\chi}}-v_{\|,o}+B_{\|,o}+\frac{1}{4}\left[3C_{\|,o}-\left(C^i_i\right)_o\right]+\frac{1}{2}\int_0^{\overbar{\chi}}\ud\ti\chi(\overbar{\chi}-\ti\chi)\frac{\ti\chi}{\overbar{\chi}}\ti\nabla_{\perp}^2\left( A -B\pp-\frac{1}{2} C\pp\right)\\
        &\quad\,+\frac{1}{2} \int_0^{\overbar{\chi}}\ud\ti\chi\left [ \ti \p^i_{\perp}B_{\perp}^i-\frac{2B_{\pp}^i}{\ti\chi} +\Per^{ij}n^k\ti\p_i C_{jk}+\frac{1}{\ti\chi}\left(C^i_i-3C\pp\right)\right ]\\
        &=\kappa_o+\kappa\,,
    \end{split} 
\end{equation}
where with $\kappa$ we indicate the integrated terms and with $\kappa_0$ the terms at the observer.
Finally the expression of $\DL$ can be written in the gauge-free setting as
\begin{equation} \begin{split}
    \Delta\ln\DL=&\,\, \int_{\overbar{V}}\ud\overbar{\x}^3\Biggl\{\WD\biggl\{- \kappa-\frac{T}{\overbar{\chi}}+\frac{1}{4}\Per^{ij}C_{ij}+\left (-\frac{1}{\overbar{\chi}\cH}+1\right)v\pp+\left (\frac{1}{\overbar{\chi}\cH}-1\right)A+2\left (-\frac{1}{\overbar{\chi}\cH}+1\right)I\\
    &+\frac{1}{\overbar{\chi}}\delta x^0_o+\left (\frac{1}{\overbar{\chi}\cH}\right)v_{\|,o}+\left (-\frac{1}{\overbar{\chi}\cH}+1\right)(\delta a_o+A_o)\\
    &-\left[B_{\|,o}+\frac{3}{4}C_{\|,o}-\frac{1}{4}\left(C^i_i\right)_o\right]\biggl\}+\left(\WD-\WV\right)\biggl\{\delta_{\rm GW}\\
    &+(5s-2)\left ( \kappa+\frac{T}{\overbar{\chi}}\right)-\left(\frac{5s}{4}-\frac{1}{2}\right)\Per^{ij}C_{ij}+\left [\frac{(5s-2)}{\overbar{\chi}\cH}-5s+b_{\rm e}-\frac{\cH'}{\cH^2}\right]v\pp\\
    &-\left [\frac{(5s-2)}{\overbar{\chi}\cH}-5s-1+b_{\rm e}-\frac{\cH'}{\cH^2}\right]A+\left [\frac{(10s-4)}{\overbar{\chi}\cH}-10s+2b_{\rm e}-2\frac{\cH'}{\cH^2}\right]I\\
    &-\frac{1}{\cH}\p\pp v\pp-\frac{1}{2\cH}C'\pp-\frac{(5s-2)}{\overbar{\chi}}\delta x^0_o+\left [-\frac{(5s-2)}{\overbar{\chi}\cH}+2-b_{\rm e}+\frac{\cH'}{\cH^2}\right]v_{\|,o}\\
    &+\left [\frac{(5s-2)}{\overbar{\chi}\cH}-5s+b_{\rm e}-\frac{\cH'}{\cH^2}\right](\delta a_o+A_o)\\
    &+(5s-2)\left[B_{\|,o}+\frac{3}{4}C_{\|,o}-\frac{1}{4}\left(C^i_i\right)_o\right]\biggl\}\Biggl\}\, ,
    \label{GR_correction_Delta_DL_generic_gauge}
\end{split} \end{equation}
where the evolution bias $b_e$ has been defined in Eq.~\eqref{def:evolution_bias} and $\delta_{\rm GW}$ is the density perturbation of GW sources and is not a gauge invariant quantity, so to define correctly the bias we need to set a gauge.
From \eqref{GR_correction_Delta_DL_generic_gauge} we can follow the procedure already described in section \ref{Angular power spectrum} (i.e. project onto spherical harmonics and go in Fourier space), then one can proceed to the evaluation of the angular power spectrum defined in Eq.~\eqref{Cl general}. In that computation we have the term $\Delta\ln\D_L$ plus the number density perturbation $\Delta_{\rm GW}$ (which enters in $\Delta\ln\DL$ thanks to the averaging process). Here we write explicitly the terms in Eq.\eqref{eq:delta_GW_as_sum_of_terms}, already in the Poisson gauge, which is the one where the numerical evaluation is performed
\begin{equation}
\begin{aligned}\label{eq:source_functions_Weighted_DL}
\left[\Delta\ln\D_L\right]_\ell^{\rm vel}=&\left(1-\frac{1}{\chi \mathcal{H}}\right) j_{\ell}^{\prime}(k\chi) \,v\left(k, \D_L\right), \\
\left[\Delta\ln\D_L\right]_\ell^{\rm lens}=&\ell(\ell+1) \int_0^{\chi} \ud \tilde\chi \frac{\chi-\tilde\chi}{\chi \tilde\chi}\Phi\left(k, \D_L\right) j_{\ell}(k \tilde{\chi})\\
\left[\Delta\ln\D_L\right]_\ell^{\rm GR}=&
\left(\frac{1}{\chi \mathcal{H}}-2\right) j_{\ell}\Phi\left(k, \D_L\right) + 2\int_0^{\chi} \ud\tilde{\chi} \frac{1}{\chi}j_{\ell}(k\tilde{\chi} )\Phi(k, \D_L) \\
&+ 2\int_0^{\chi} \ud\tilde{\chi}\left(\frac{1}{\chi \mathcal{H}}-1\right)j_{\ell}(k \tilde{\chi})\Phi^{\prime}(k, \D_L) 
\\
\Delta_{\ell}^{\mathrm{den}}(k, \D_L)= & b_{\rm GW} \,\delta\left(k, \D_L\right) j_{\ell}(k\chi) \\ \Delta_{\ell}^{\mathrm{vel}}(k, \D_L)= & \Delta_{\ell}^{\mathrm{rsd}}(k, \D_L)+\Delta_{\ell}^{\mathrm{D1}}(k, \D_L)+\Delta_{\ell}^{\mathrm{D2}}(k, \D_L), \\ \Delta_{\ell}^{\mathrm{rsd}}(k, \D_L)= & -\frac{k}{\mathcal{H}} j_{\ell}^{\prime \prime}(k\chi) v\left(k, \D_L\right), \\ 
\Delta_{\ell}^{\mathrm{D1}}(k, \D_L)= & \left(-\frac{\mathcal{H}^{\prime}}{\mathcal{H}^2}+\frac{5s-2}{\chi \mathcal{H}}-5 s+b_{\rm e}\right) j_{\ell}^{\prime}(k\chi) \,v\left(k, \D_L\right),  \\
\Delta_{\ell}^{\mathrm{D2}}(k, \D_L)= & -\left(b_{\rm e}-3\right) \frac{\mathcal{H}}{k} j_{\ell}(k\chi)\, v\left(k, \D_L\right),  \\
\Delta_{\ell}^{\mathrm{lens}}(k, \D_L)= & \ell(\ell+1) (2-5 s) \int_0^{\chi} \ud \tilde\chi \frac{\chi-\tilde\chi}{\chi \tilde\chi}\Phi\left(k, \D_L\right) j_{\ell}(k \tilde{\chi}), \\ 
\Delta_{\ell}^{\mathrm{GR}}(k, \D_L)= & \sum_{i=1}^5\Delta_{\ell}^{\mathrm{GR\,i}}(k, \D_L)\\
\Delta_{\ell}^{\mathrm{GR\,1}}(k, \D_L)=&\left(\frac{\mathcal{H}^{\prime}}{\mathcal{H}^2}+\frac{2-5 s}{\chi \mathcal{H}}+5 s-b_{\rm e}+1\right) j_{\ell}(k\chi)\Phi\left(k, \D_L\right)\\
\Delta_{\ell}^{\mathrm{GR\,2}}(k, \D_L)=&\left(-2+5 s\right) j_{\ell}(k\chi)\Phi\left(k, \D_L\right)\\
\Delta_{\ell}^{\mathrm{GR\,3}}(k, \D_L)=&\mathcal{H}^{-1} j_{\ell}(k\chi)\Phi^{\prime}\left(k, \D_L\right)  \\ 
\Delta_{\ell}^{\mathrm{GR\,4}}(k, \D_L)=& 2\int_0^{\chi} \ud\tilde{\chi} \frac{2-5 s}{\chi}j_{\ell}(k\tilde{\chi} )\Phi(k, \D_L) , \\ 
\Delta_{\ell}^{\mathrm{GR\,5}}(k, \D_L)=& 2\int_0^{\chi} \ud\tilde{\chi}\left(\frac{\mathcal{H}^{\prime}}{\mathcal{H}^2}+\frac{2-5 s}{\chi \mathcal{H}}+5 s-b_{\rm e}\right)_{\chi}j_{\ell}(k \tilde{\chi})\Phi^{\prime}(k, \D_L)  .
\end{aligned}
\end{equation}
In these equations we replaced $\overbar{\chi}$ with $\chi$ to simplify the notation. 
One can see that the terms $\Delta\ln\D_L^\alpha$ correspond to only the contributions $\Delta_{\rm GW}^\alpha$ proportional to the magnification bias $s$, fixing $s=-0.2$.
To calculate numerically the angular power spectrum, we used a modified version of \texttt{Multi\_class}~\cite{Bellomo:2020pnw,Bernal:2020pwq}, an extension of CLASSgal~\cite{DiDio:2013bqa}, which gives the possibility of computing the angular power spectrum of the cross-correlation between different tracers.
The scale-free bias $b_{\rm GW}$ is defined in terms of the density contrasts of GWs and matter in the synchronous-comoving gauge (SC) by
\begin{equation}
    \delta^{\rm SC}_{\rm GW} = b_{\rm GW}\delta_m^{\rm SC} \, .
\end{equation}
In the literature, $\delta_m^{\rm SC}$ is often indicated as $D$, see, e.g.,~\cite{DiDio:2013bqa}. In order to express the density perturbation of the GWs in the Poisson gauge (P) as a function of the scale invariant bias we follow~\cite{Jeong:2011as} and we use
\begin{equation}
   \delta_{\rm GW}^{\rm P}= b_{\rm GW} \,\left(\delta^{\rm P}_m-3\cH v\right) - \left(b_{\rm e}-3\right)\cH v\, ,
    \label{eq:delta_GW_Poisson_delta_m_Poisson} 
\end{equation}
where 
\begin{equation}\begin{split}
    \delta^{\rm P}_m-3\cH v = \delta^{\rm SC}_m \, ,
    \label{eq:delta_m_Poisson_SC}
\end{split}\end{equation}
where the velocity potential is, $v_i=\partial_i v$. 

\section{Window function}

\begin{figure}[h!]
    \includegraphics[width=\linewidth]{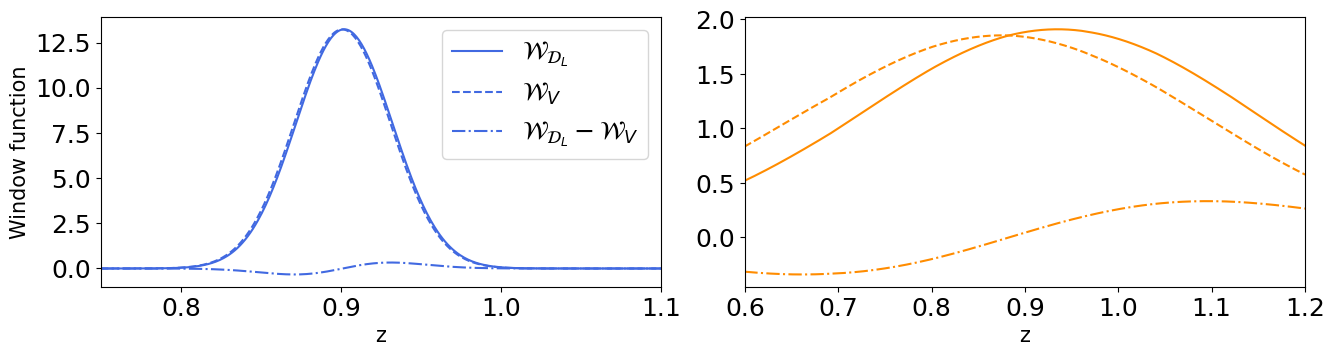}
    \caption{Plot of $\WD$, $\WV$ and the difference of the two for BBH (left) and BNS (right). In  this Figure we consider $\D_L^{\rm bin}=0.9$ for both sources. We can see that the worse LD precision of BNS strongly impact on the bin shape.}
    \label{fig:window}
\end{figure}

In this section we present the window functions introduced in Eqs.\eqref{def:window_functions_V_DL}. In the left and right panels of Figure~\ref{fig:window}, we plot the window functions $\WD$ and $\WV$ for BBH and BNS respectively. In the plots we show also the difference between the two window functions, which multiplies $\Delta_{\rm GW}$ in Eq.~\eqref{eq:decomposition_W_DL}. We notice that the difference between the window functions is small, in particular in the BBH case, because of the smaller bin size (the average precision expected at that LD), therefore the contribution to $\Delta\ln\DL$ given by $\Delta_{\rm GW}$ is suppressed w.r.t. to the one induced by $\Delta\ln\D_L$. 

\section{Binning}
\label{sec:Binning}

We select 40 Gaussian bins centered at luminosity distances which corresponds to redshifts between $z_{\rm min}=0.1$ and $z_{\rm max}=4.0$. The bins are chosen in such a way that the width of the bin is equal to the instrumental uncertainty on the estimate of the LD of the sources inside the bin. The binning adopted in this work is not the only possible choice, but the analysis of which $z_{\rm mean}$ and $\sigma$ maximize the SNR of the signal is beyond the purposes of this work. For this reason, we have fixed the widths of the bins to the average instrumental error on the LD of the sources moreover this allows to cover all the redshift range within the $2.5\sigma$ of the Gaussian window functions. We have checked that by reducing the number of bins and increasing the widths the SNR of the intrinsic anisotropies changes roughly by a factor $\sqrt{N_{\rm bin}}$, thus our findings are not too much affected by the binning.

\begin{table}[H]
\begin{center}
\begin{tabular}{|l|l|l|l|l|}
\hline
$z_{\rm mean}$ & $\sigma_{\rm ETCE}^{\rm BBH}$ & $\sigma_{\rm BBO}^{\rm BBH}$ & $\sigma_{\rm ETCE}^{\rm BNS} $& $\sigma_{\rm BBO}^{\rm BNS}$\\
\hline
0.1  &  0.001  &  0.0001  &  0.005  &  0.0003\\ 
\hline
0.2  &  0.003  &  0.0002  &  0.015  &  0.0005\\ 
\hline
0.3  &  0.004  &  0.0003  &  0.025  &  0.001\\ 
\hline
0.4  &  0.007  &  0.0004  &  0.042  &  0.0019\\ 
\hline
0.5  &  0.011  &  0.0006  &  0.066  &  0.0029\\ 
\hline
0.6  &  0.015  &  0.0008  &  0.097  &  0.004\\ 
\hline
0.7  &  0.02  &  0.0011  &  0.143  &  0.0054\\ 
\hline
0.8  &  0.025  &  0.0014  &  0.198  &  0.0069\\ 
\hline
0.9  &  0.03  &  0.0017  &  0.255  &  0.0085\\ 
\hline
1.0  &  0.036  &  0.0021  &  0.319  &  0.0103\\ 
\hline
1.1  &  0.042  &  0.0025  &  0.382  &  0.0122\\ 
\hline
1.2  &  0.048  &  0.0029  &  0.437  &  0.0143\\ 
\hline
1.3  &  0.056  &  0.0034  &  0.488  &  0.0164\\ 
\hline
1.4  &  0.062  &  0.0038  &  0.54  &  0.0187\\ 
\hline
1.5  &  0.071  &  0.0044  &  0.591  &  0.0211\\ 
\hline
1.6  &  0.082  &  0.0049  &  0.643  &  0.0236\\ 
\hline
1.7  &  0.09  &  0.0054  &  0.693  &  0.0262\\ 
\hline
1.8  &  0.098  &  0.006  &  0.744  &  0.029\\ 
\hline
1.9  &  0.103  &  0.0066  &  0.795  &  0.0318\\ 
\hline
2.0  &  0.107  &  0.0072  &  0.846  &  0.0347\\ 
\hline
\end{tabular}
\quad
\begin{tabular}{|l|l|l|l|l|}
\hline
$z_{mean}$ & $\sigma_{\rm ETCE}^{\rm BBH}$ & $\sigma_{\rm BBO}^{\rm BBH}$ & $\sigma_{\rm ETCE}^{\rm BNS} $& $\sigma_{\rm BBO}^{\rm BNS}$\\

\hline
2.1  &  0.112  &  0.008  &  \quad \textbackslash  &  0.038\\ 
\hline
2.2  &  0.116  &  0.008  &  \quad \textbackslash  &  0.041\\ 
\hline
2.3  &  0.121  &  0.009  &  \quad \textbackslash  &  0.044\\ 
\hline
2.4  &  0.125  &  0.01  &  \quad \textbackslash  &  0.047\\ 
\hline
2.5  &  0.13  &  0.01  &  \quad \textbackslash  &  0.051\\ 
\hline
2.6  &  0.134  &  0.011  &  \quad \textbackslash  &  0.054\\ 
\hline
2.7  &  0.139  &  0.012  &  \quad \textbackslash  &  0.058\\ 
\hline
2.8  &  0.143  &  0.013  &  \quad \textbackslash  &  0.061\\ 
\hline
2.9  &  0.148  &  0.013  &  \quad \textbackslash  &  0.065\\ 
\hline
3.0  &  0.152  &  0.014  &  \quad \textbackslash  &  0.069\\ 
\hline
3.1  &  0.156  &  0.015  &  \quad \textbackslash  &  0.073\\ 
\hline
3.2  &  0.159  &  0.016  &  \quad \textbackslash  &  0.076\\ 
\hline
3.3  &  0.163  &  0.016  &  \quad \textbackslash  &  0.08\\ 
\hline
3.4  &  0.166  &  0.017  &  \quad \textbackslash  &  0.084\\ 
\hline
3.5  &  0.17  &  0.018  &  \quad \textbackslash  &  0.088\\ 
\hline
3.6  &  0.174  &  0.019  &  \quad \textbackslash  &  0.092\\ 
\hline
3.7  &  0.177  &  0.02  &  \quad \textbackslash  &  0.096\\ 
\hline
3.8  &  0.181  &  0.021  &  \quad \textbackslash  &  0.101\\ 
\hline
3.9  &  0.184  &  0.021  &  \quad \textbackslash  &  0.105\\ 
\hline
4.0  &  0.188  &  0.022  &  \quad \textbackslash  &  0.109\\ 
\hline

\end{tabular}
\end{center}
\end{table}

\section{Velocity contribution }
\label{app:vel}

\begin{figure}[h!]
    \includegraphics[width=\linewidth]{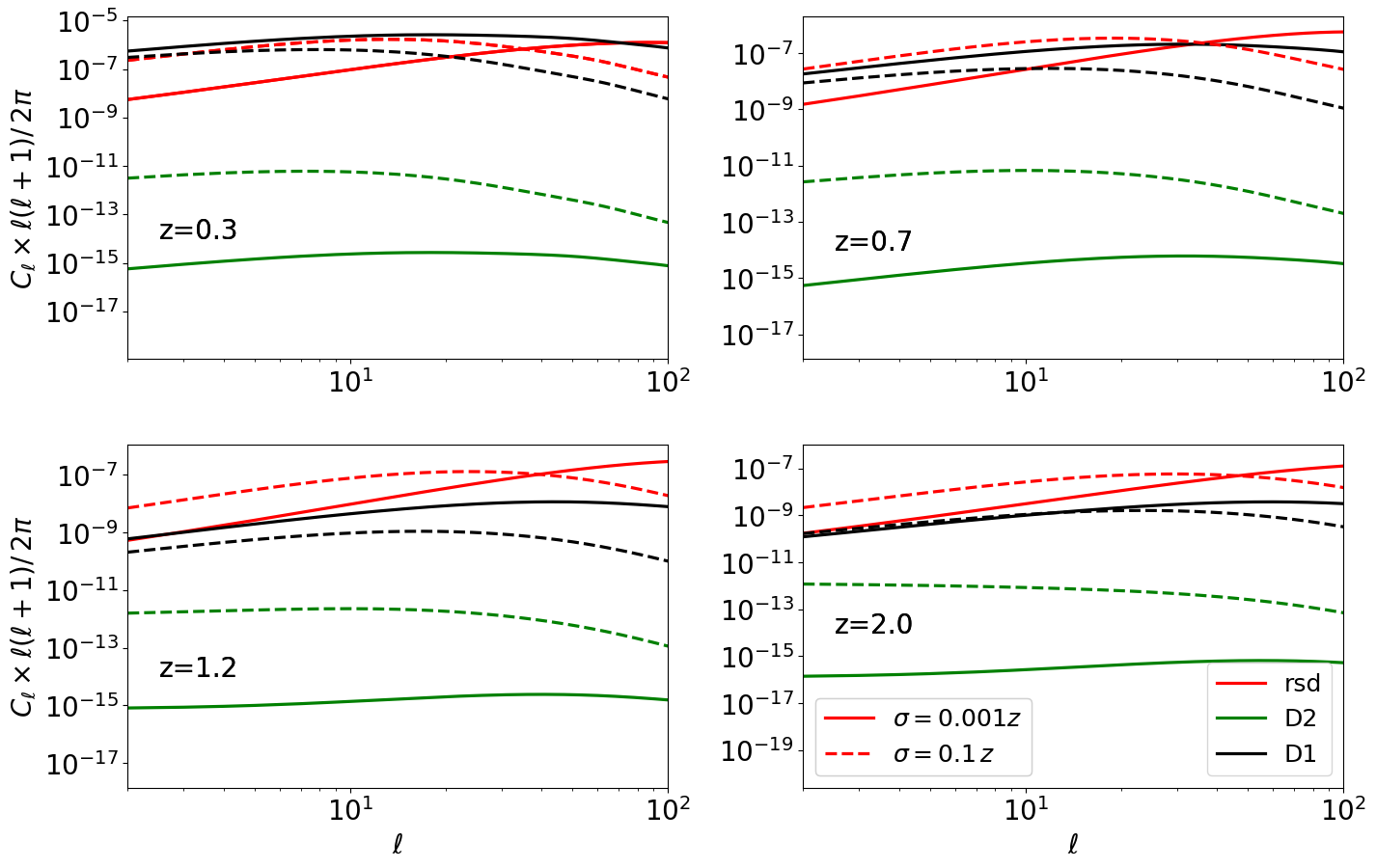}
    \caption{Plot of the three contributions of the peculiar velocity term. They are indicated with rsd (the Kaiser term), D1 and D2 (the Doppler terms) and they are represented for two different bin widths. One can see that for low redshift and thin bins the D1 contribution dominates, while rsd dominates in the other cases.}
    \label{fig:width_vel}
\end{figure}

In this appendix we explain in more details the origin of the behaviour of the peculiar velocity term when changing the width of the bin. We see (from Figure \ref{fig:Cl_GR_corrections}) that at low redshift the peculiar velocity term increases with the bin width while the opposite happens at higher redshifts.
The reason for this behaviour can be seen in Figure \ref{fig:width_vel}. In this Figure we plot the three contributions of velocity term to $\Delta\ln\DL$: the Kaiser term (indicated with rsd) plus two Doppler terms (D1 and D2). When we refer to these terms in the plot we consider them already with the window functions. These terms are inside the definition of $\Delta\ln\DL$, i.e. they already contain the window functions w.r.t. the velocity terms defined in \ref{eq:source_functions_Weighted_DL}. The rsd and the D2 term are present only in $\Delta_{\rm GW}$ while D1 is present in $\Delta_{\rm GW}$ and in $\Delta\ln\D_L$ (one can recognize the same dependence on the spherical Bessel derivative in $\Delta^{\rm D1}_{\ell}$ and in $[\Delta\ln\D_L]_\ell^{\rm vel}$). We see that rsd and D2 decrease with thinner bins while D1 does the opposite. This happens because the D1 term is present in both $\Delta_{\rm GW}$ and $\Delta\ln\D_L$, and $\Delta\ln\D_L$ increases with thinner bins, counter balancing the D1 term present in $\Delta_{\rm GW}$. 
The fact that at low redshift and thin bins D1 dominates while at high redshift rsd dominates give the changing behaviour of the velocity term of $\Delta\ln\DL$ with the bin width.
One can also note a scale dependence in how the different contributions are affected by the change in width. This can be understood from the derivative of the Bessel function, since they are evaluated as $\approx j_\ell(k)/k$, leading to the scale dependence.

\section{Derivation of the Kinetic Dipole}
\label{app:derivation of the kinetic dipole}

In Eq.~\eqref{eq:total formula pois} we recognize several terms proportional to the velocity of the observer which contribute to the anisotropies of $\DL$,  
\begin{equation} \begin{split}
    (\Delta\ln\DL)_{o}&=\left[\WD\frac{1}{\overbar{\chi}\cH}+\left(\WD- \WV\right) \left(-\frac{5s-2}{\overbar{\chi}\cH}+2-b_{\rm e}+\frac{\cH'}{\cH^2}\right)\right]v_{\|,o}\,.
\end{split} \end{equation}
By using the relation between the Legendre polynomials and the spherical harmonics
\begin{equation} \begin{split}
    \label{identity daniele}           \mathcal{P}_\ell\left(\hat{\n}\cdot\hat{\n}^\prime\right)=\frac{4\pi}{2\ell+1}\sum_{m^\prime=-\ell}^{+\ell}Y^*_{\ell m^\prime}(\hat{\n})Y_{\ell m^\prime}(\hat{\n}^\prime)\,,
\end{split} \end{equation}
and 
\begin{equation}
    v_{\parallel,o} = \boldsymbol{v}_0\cdot \hat{\n} = v_o\mathcal{P}_1\left(\hat{\boldsymbol{v}}_0\cdot \hat{\n}\right) \, .
\end{equation}
it is possible to show that the velocity of the observer induces a dipole corrections to the average LD,
\begin{equation}
    \left[\Delta\ln\mathscr{D}_{L,o}\right]_{\ell m}= \frac{4\pi}{3}\,\delta_{1\ell}  v_o Y_{1 m}^*(\hat{\boldsymbol{v}}_o)\left[\WD\frac{1}{\overbar{\chi}\cH}+\left(\WD- \WV\right) \left(-\frac{5s-2}{\overbar{\chi}\cH}+2-b_{\rm e}+\frac{\cH'}{\cH^2}\right)\right]\, , 
\end{equation}
where $\left[\Delta\ln\mathscr{D}_{L,o}\right]_{\ell m}$ represents the observer correction to the weighted LD in harmonic space.
This calculation is done according to the normalization of the spherical harmonics
\begin{equation} \begin{split}
   \sum_{m^\prime=-1}^{+1}|Y_{1 m^\prime}(\hat{\n})|^2=\frac{3}{4\pi}\,.
\end{split} \end{equation}
Then the covariance of the dipole between two different bins is 
\begin{equation} \begin{split}
    C_{\DL}^{\rm KD}\left(\D_{L,i}^{\rm b},\D_{L,j}^{\rm b}\right)&=\frac{4\pi }{9}v_{o}^2\delta_{\ell\,1}\int\ud \D_L \int \ud \tilde{\D}_L \\
    &\times\left[\WD\frac{1}{\overbar{\chi}\cH}+\left(\WD- \WV\right) \left(-\frac{5s-2}{\overbar{\chi}\cH}+2-b_{\rm e}+\frac{\cH'}{\cH^2}\right)\right]\left(\D_L,\D_{L,i}^{\rm b}\right)\\
    &\times\left[\WD\frac{1}{\overbar{\chi}\cH}+\left(\WD- \WV\right) \left(-\frac{5s-2}{\overbar{\chi}\cH}+2-b_{\rm e}+\frac{\cH'}{\cH^2}\right)\right]\left(\tilde{\D}_L,\D_{L,j}^{\rm b}\right) \, ,
\end{split} \end{equation}
where the functions inside the square brackets are evaluated in the $i-\rm th$ and $j-th$ bins respectively and the variable to span the bin is $\D_{L}$.

\section{Shot Noise}
\label{app: shot noise}
\begin{figure}[H]
    \centering
    \includegraphics[width=0.6\linewidth]{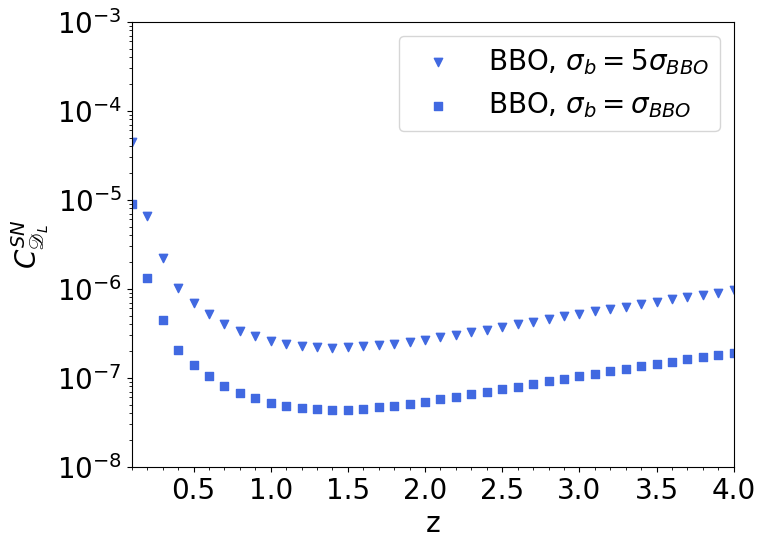}
    \caption{In this Figure we show the shot noise for two different bin width, here called $\sigma_{\rm b}$, which are a multipole of the instrumental uncertainty $\sigma_{\rm BBO}$ and we can see that the thinner bins (squares) have a smaller shot noise than the larger bins (triangles).}
    \label{fig:shot_noise_bins}
\end{figure}

\begin{figure}[H]
    \centering
    \includegraphics[width=0.7\linewidth]{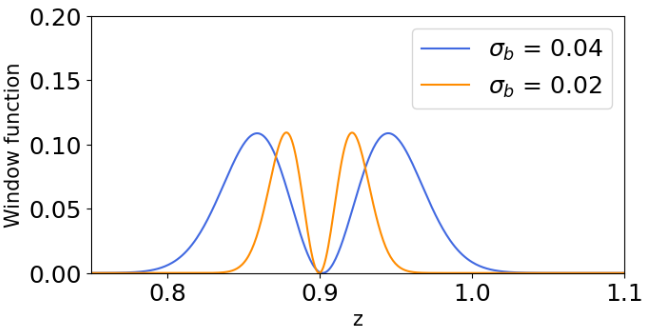}
    \caption{In this plot we show the term $\left(\mathcal{W}_{\D_L} -\mathcal{W}_{V}\right)^2$ for two bin width choices. We can see that the area under the curve is larger in the case of a larger bin width, leading to the increase of the shot noise. }
    \label{fig:windows}
\end{figure}

The average number of GW events expected in an infinitesimal comoving volume $\ud V_{\rm c}$ are 
\begin{equation}
    \begin{split}
        \left\langle n_{\rm GW}\left(\D_L,\hat{\n}\right)\ud V_{\rm c} \right \rangle = & \overbar{n}_{\rm GW}\ud V_{\rm c}\, .
    \end{split}
\end{equation}
The brackets here indicate an average over many realizations of the shot noise. Thanks to the ergodic theorem, this average is equivalent to the average over the solid angle at a given $\D_L$. The covariance of the shot noise fluctuations in the number of objects is 
\begin{equation} \begin{split}
        \label{covariance_dNdV} 
        C^{\rm SN}\left(\D_{L,1}\right)=&\left \langle \left[n_{\rm GW}(\D_{L,1},\hat{\n}_1)-\overbar{n}_{\rm GW}(\D_{L,1})\right]\ud V_{c,1} \left[n_{\rm GW}(\D_{L,2},\hat{\n}_2)-\overbar{n}_{\rm GW}(\D_{L,2})\right]\ud V_{c,2}\right\rangle \\
        =&\delta(\D_{L,1}-\D_{L,2})\delta(\hat{\n}_1-\hat{\n}_2) \,\overbar{n}_{\rm GW}(\D_{L,1})\, \ud V_{c,1} \, \ud V_{c,2}\, ,
\end{split} \end{equation}
where we have used the fact that Poisson fluctuations in different regions in the sky are independent. To simplify the notation, as we did in the main text, we write the number of objects in terms of the density contrast due to the shot noise
\begin{equation}
    n_{\rm GW}(\mathcal{D}_L,\hat{\n}) = \overbar{n}_{\rm GW}(\mathcal{D}_L)\left[1+\delta_{\rm SN}(\mathcal{D}_L,\hat{\n})\right] \, ,
    \label{eq:exp_N_GW_delta_SN}
\end{equation}
where the fluctuation of the density contrast $\delta_{\rm SN}$ is a Gaussian random variable with mean and covariance given by 
\begin{equation}
    \begin{split}
   \left\langle  \delta_{\rm SN}(\mathcal{D}_L,\hat{\n})\right \rangle &=0  \, ,  \\
    \left\langle \prod_{i=1}^2 \delta_{\rm SN}(\mathcal{D}_{L,i},\hat{\n}_i)\,\overbar{n}_{\rm GW}(\mathcal{D}_{L,i})\,\ud V_{c,i}\right \rangle &=
    \overbar{n}_{\rm GW}\ud V_{c,1} \ud V_{c,2}\,  \delta(\D_{L,1}-\D_{L,2})\delta(\hat{\n}_1-\hat{\n}_2)   \, .    
    \label{eq:delta_sn_correlators}
    \end{split}
\end{equation}
In the redshift bins considered, the shot noise is small, $\delta_{\rm SN}\lesssim 10^{-2}$, therefore we can expand $\DL$ up to terms linear in the shot noise. 
\begin{equation} \begin{split}
    \mathscr{D}_L&=\frac{\int\ud\overbar{\x}^3\, W_{\rm b}\BD_L\overbar{n}^{\rm ph}_{\rm GW}\left(1+\delta_{\rm SN}\right)}{\int\ud\overbar{\x}^3\, W_{\rm b}\, \overbar{n}^{\rm ph}_{\rm GW}\left(1+\delta_{\rm SN}\right)}\,,
\end{split} \end{equation}
thus we obtain
\begin{equation}
    \begin{split}
        \DL= &  \BDL \left(1 + \int \ud\overbar{\x}^3 \left(\WD -\WV \right) \delta_{SN}\right) \, , 
    \end{split}
\end{equation}
therefore the perturbation of the shot noise $\sigma_{\rm SN}$ defined in Eq.~\eqref{eq:def_bar_DL_discrete} is just
\begin{equation}
    \ln \sigma_{\rm SN} = \int \ud\overbar{\x}^3 \left(\WD -\WV \right) \delta_{SN} \, .
\end{equation} 
We write the covariance associated to the shot noise as
\begin{equation}
\begin{split}
    C^{\rm SN}_{\DL}&={\rm cov}\left[\ln \sigma_{\rm SN}(\D_{L,1},\hat{\n}_1),\ln \sigma_{\rm SN}(\D_{L,2},\hat{\n}_2)\right]=\\
    &\left\langle\int \ud \overbar{\x}^3_1 \ud \overbar{\x}^3_2 \left(\mathcal{W}_{\D_L 1} -\mathcal{W}_{V 1} \right) \delta_{SN}(\D_{L,1},\hat{\n}_1) \left(\mathcal{W}_{\D_L 2} -\mathcal{W}_{V 2} \right) \delta_{SN}(\D_{L,2},\hat{\n}_2)\right\rangle \\
    &=\int \ud \overbar{\x}^3_1\ud \overbar{\x}^3_2\,\overbar{n}_{\rm GW1} \delta(\D_{L1}-\D_{L2})\delta(\hat{\n}_1-\hat{\n}_2)\frac{\mathcal{W}_{\D_L 1} -\mathcal{W}_{V 1}}{\overbar{n}_{\rm GW1}}\frac{\mathcal{W}_{\D_L 2} -\mathcal{W}_{V 2}}{\overbar{n}_{\rm GW2}}\\
    &=\int  \frac{\ud \overbar{\x}^3}{\overbar{n}_{\rm GW}}\left(\mathcal{W}_{\D_L 1} -\mathcal{W}_{V 1}\right)\left(\mathcal{W}_{\D_L 2} -\mathcal{W}_{V 2}\right) \, .
    \label{eq:sn_different_bins}
\end{split}
\end{equation}
We stress that in the previous equation the window functions are evaluated in the same point $\left(\D_L,\hat{\n}\right)$, but they could be centered in different bins, $\left(\D_{L,1/2}^{\rm b},\hat{\n}^{\rm b}_{1/2}\right)$.
Considering the autocorrelation of the shot noise (i.e. the diagonal of the covariance matrix) we have

\begin{equation}
\begin{split}
    C^{\rm SN}_{\DL}&=\int  \frac{\ud \overbar{\x}^3}{\overbar{n}_{\rm GW}}\left(\mathcal{W}_{\D_L} -\mathcal{W}_{V}\right)^2 \, .
    \label{eq:sn_same_bin}
\end{split}
\end{equation}
As remarked in Section \ref{sec:shot_noise}, we show in Figure \ref{fig:shot_noise_bins} that if we increase the bin width, the shot noise increases. 
This is driven by the term $\left(\mathcal{W}_{\D_L} -\mathcal{W}_{V}\right)^2$ in the definition of the shot noise, as can be seen from Figure \ref{fig:windows}

\section{Instrumental error}\label{app:instrumental_error}

The estimator of the average LD is defined by
\begin{equation}
    \DL\left(\D_L^{\rm b},\bn^{\rm b}\right) \equiv \frac{1}{N_{\rm GW}^{\rm b}}\sum_{i} W_{\rm b}\left(\D_{L (i)},\bn_{(i)}\right) \D_{L(i)}\, ,
\end{equation}
where we have considered the Gaussian window, Eq.~\eqref{eq:def_W_c}, and the number of objects in the bin is 
\begin{equation}
    N_{\rm GW}^{\rm b}\left(\D_L^{\rm b},\bn^{\rm b}\right) \equiv \sum_{i} W_{\rm b}\left(\D_{L (i)},\bn_{(i)}\right) \, .
\end{equation}
The estimated LDs and angular positions of the individual sources $\D_{L(i)}$ and $\bn_{(i)}$ deviates from the background values $\BD_{L(i)}$ and $\hat{\overbar{\textbf{n}}}_{(i)}$ because of GR corrections, the kinetic dipole and instrumental noise,
\begin{equation}
    \begin{split}
        \D_{L(i)} = & \BD_{L(i)}\left[1+\Delta\ln\D_{L}\left(\BD_{L (i)},\hat{\overbar{\textbf{n}}}_{(i)}\right)\right]+\delta\mathcal{D}_L^{\rm instr}\left(\BD_{L (i)},\hat{\overbar{\textbf{n}}}_{(i)}\right) \, , \\
        \bn_{(i)} =& \hat{\overbar{\textbf{n}}}_{(i)}\left[1+\Delta\ln\bn\left(\BD_{L (i)},\hat{\overbar{\textbf{n}}}_{(i)}\right)\right]+\delta\bn^{\rm instr}\left(\BD_{L (i)},\hat{\overbar{\textbf{n}}}_{(i)}\right)\, .
    \end{split}
\end{equation}
The uncertainties on the estimator of the LD due to the GR and kinetic corrections, encoded in $\Delta \ln \D_L$, have already been computed in Section~\ref{The Weighted Luminosity Distance}, thus in this section we focus just on the covariance induced by the instrumental noise. For two sources $(i)$ and $(j)$, we quantify the total instrumental noise according to  
\begin{equation}
\begin{split}
    {\rm cov}\left[\delta\mathcal{D}_L^{\rm instr}\left(\BD_{L (i)},\hat{\overbar{\textbf{n}}}_{(i)}\right),\delta\mathcal{D}_L^{\rm instr}\left(\BD_{L (j)},\hat{\overbar{\textbf{n}}}_{(j)}\right)\right] \equiv & \delta_{(i)(j)}   \sigma^2_{\mathcal{D}_L}\left(\BD_{L (i)},\hat{\overbar{\textbf{n}}}_{(i)}\right) \, , \\
    {\rm cov}\left[\delta\bn^{\rm instr}\left(\BD_{L (i)},\hat{\overbar{\textbf{n}}}_{(i)}\right),\delta\bn^{\rm instr}\left(\BD_{L (j)},\hat{\overbar{\textbf{n}}}_{(j)}\right)\right] \equiv& \delta_{(i)(j)}\sigma^2_{\hat{\n}}\left(\BD_{L (i)},\hat{\overbar{\textbf{n}}}_{(i)}\right) \, , \\
     \label{eq:instrumental_covariances}
\end{split}
\end{equation}
where we have assumed that the instrumental errors between two different sources are uncorrelated. We have not considered the covariance between the error on the LD and the angular resolution, since it is give a subdominant contribution to the error on the estimate of the average LD. The errors on the LD and the angular resolution of the individual sources have been plotted in Figure~\ref{fig:sn_instrn}, while details about their computation can be found in Appendix~\ref{app:population_source_and_instrumental_noise}.

The impact of $\sigma_{\D_L}$ and $\sigma_{\hat{\n}}$ on $\DL$ can be found by using the standard propagation of error. More explicitly, the lack of knowledge on $\D_{L(i)}$ and $\hat{\n}_{(i)}$ corresponds to an intrinsic uncertainty on the window function of the bin used. Since we have accounted for the uncertainties due to the angular resolution of the sources by convolving the window function of the bin with the beam function defined in Eq.~\eqref{def:cut_ell_max_instr}, we consider in this section only the impact of $\sigma_{\D_L}$. The uncertainty on the LD of the individual sources propagates in the estimate of the objects inside the bin according to
\begin{equation}
    N_{\rm GW}^{\rm b}\left(\D_L^{\rm b},\bn^{\rm b}\right) \equiv \overbar{N}_{\rm GW}^{\rm b}\left(\D_L^{\rm b},\bn^{\rm b}\right)+\sum_{i}
        \partial_{\BD_{L(i)}} 
        W_{\rm b}\left(\BD_{L (i)},\hat{\overbar{\textbf{n}}}_{(i)}\right) 
        \delta\D_L^{\rm instr}\left(\BD_{L (i)},\hat{\overbar{\textbf{n}}}_{(i)}\right)      \, ,    
\end{equation}
where we have defined
\begin{equation}
    \overbar{N}_{\rm GW}^{\rm b}\left(\D_L^{\rm b},\bn^{\rm b}\right) \equiv \sum_{i} W_{\rm b}\left(\BD_{L (i)},\hat{\overbar{\textbf{n}}}_{(i)}\right) \, .
\end{equation}
It is possible to truncate the Taylor expansion at first order, because the relative errors are of the order of $10\%$, therefore quadratic terms will be suppressed by at least one order of magnitude. The propagation of the error on the estimator of the average LD gives
\begin{equation}
    \DL \equiv \BDL+\frac{\sum_{i} 
        \partial_{\BD_{L(i)}} 
        \left\{W_{\rm b}\, \left[\BD_{L (i)}-\BDL\right]\right\} 
        \delta\D_L^{\rm instr} }{{\overbar{N}_{\rm GW}^{\rm b}}} \, ,
\end{equation}
where we have omitted the dependences on $\D_L^{\rm b}$, $\bn^{\rm b}$, $\BD_{L (i)}$, $\hat{\overbar{\textbf{n}}}_{(i)}$ for simplicity. The covariance matrix on the estimator of the LD between two bins $\alpha$ and $\beta$ will be then
\begin{equation}
    \begin{split}
    C^{\rm instr}_{\DL}\equiv & {\rm cov}\left[\DL\left(\D_L^{\rm b-\alpha},\bn^{\rm b-\alpha}\right),\DL\left(\D_L^{\rm b-\beta},\bn^{\rm b-\beta}\right)\right]= \\
    = & \frac{\sum_i \partial_{\BD_{L(i)}} 
        \left\{W_{\rm b}^\alpha\, \left[\BD_{L (i)}-\BDL^\alpha\right]\right\}   
        \sigma^2_{\mathcal{D}_L} 
        \partial_{\BD_{L(i)}}
        \left\{W_{\rm b}^\beta\, \left[\BD_{L (i)}-\BDL^\beta\right]\right\}  }{\overbar{N}_{\rm GW}^{\rm b-\alpha}\overbar{N}_{\rm GW}^{\rm b-\beta}}
    \end{split} \, .
\end{equation}
It is straightforward to check that the continuous limit of this expression is 
\begin{equation}
    \begin{split}
    C^{\rm instr}_{\DL} = \frac{1}{{\overbar{N}_{\rm GW}^{\rm b-\alpha}\overbar{N}_{\rm GW}^{\rm b-\beta}}}\int &\ud\overbar{\x}^3\, \overbar{n}_{\rm GW} \\
        &\times
        \partial_{\mathcal{D}_L} 
        \left(W_{\rm b}^\alpha \,  \mathcal{D}_L-\BDL^\alpha W_{\rm b}^\alpha\right)        
        \sigma^2_{\mathcal{D}_L}
        \partial_{\mathcal{D}_L} 
        \left(W_{\rm b}^\beta \,  \mathcal{D}_L-\BDL^\beta W_{\rm b}^\beta\right)\delta_{\alpha\beta} \, ,
    \end{split}
    \label{eq:cov_instr_DL_DL_two_bins}
\end{equation}
where we have assumed that the bins do not overlap in the sky and therefore a Kronecker delta in the bins appears. 
This expression is consistent with the results obtained in~\cite{Namikawa:2015prh}.

\section{SNR of the single components} \label{SNR_components}

\begin{figure}[h!]
    \centering
    \includegraphics[width=\linewidth]{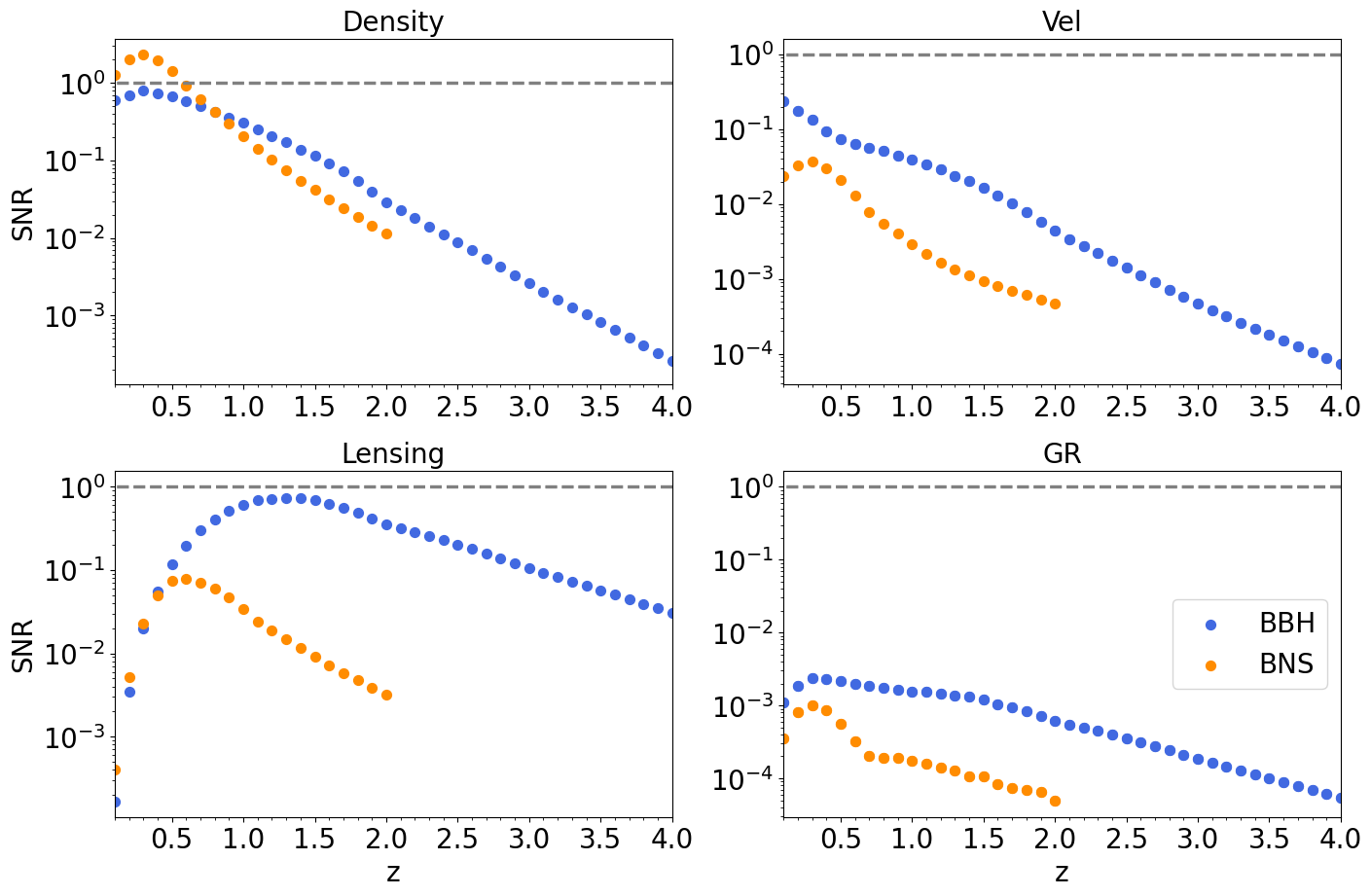}
    \caption{Plot of the SNR of the density (upper left), velocity (upper right), lensing (lower left) and GR (lower right) corrections to $\ln\DL$ for ET+2CE for $T_{\rm obs}=10\,\rm yrs$. In dashed grey we highlighted SNR=1. This plot shows the SNR per bin; each point indicates the SNR one would have when taking a bin centered at the corresponding redshift.
    }
    \label{fig_SNR_map_ET2CE}
\end{figure}

\begin{figure}[t!]
\includegraphics[width=\linewidth]{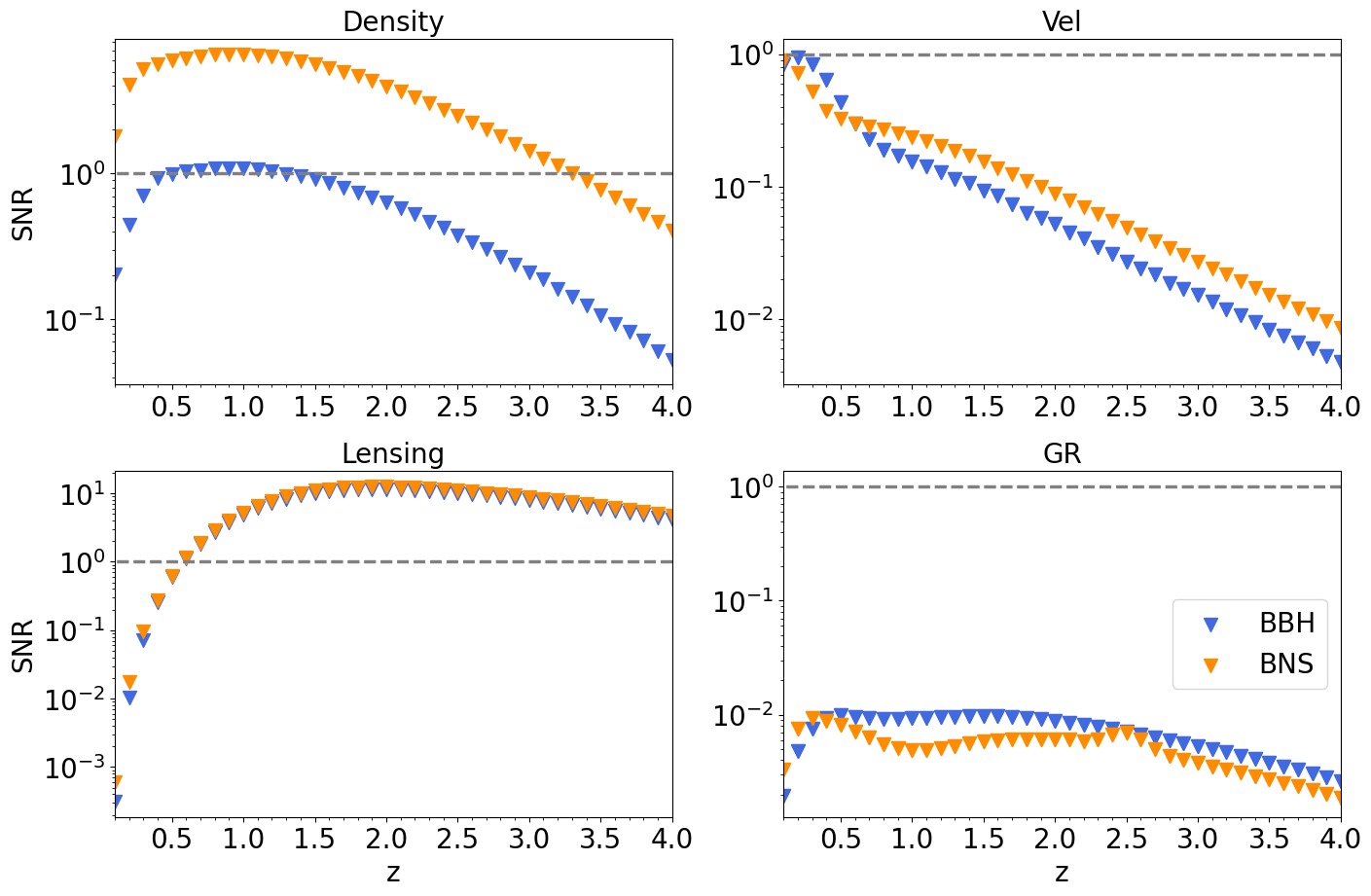}
    \caption{As in the previous Figure, but for BBO/DECIGO. We see that the density contribution has a large signal in the BNS case and there is a clear lensing signal from both probes.}
    \label{fig:SNR_Cls_BBO}
\end{figure}

In this appendix we compute the SNR of the single contributions (density, vel, lensing, GR) in Figure \ref{fig_SNR_map_ET2CE}-\ref{fig:SNR_Cls_BBO}. These plots integrate the information of Figure \ref{fig:ratio_ET2CE}-\ref{fig:ratio_BBO} 
with the different noise sources. 
We define the SNR of the angular power spectrum for a single contribution $\alpha$ (where $\alpha=$ lensing, vel, density, GR) as
\begin{equation}
    {\rm SNR_\alpha}  \equiv  \left[\sum_{\ell}\frac{(2\ell+1) B_\ell^2C_{\ell\,\alpha}^2}{2\left(C_\ell+C_{\DL}^{\rm SN}+C_{\DL}^{\rm instr}+C_{\DL}^{\rm KD}\right)^2}\right]^{1/2} \, ,
    \label{eq:SNR_pseudo_Cl}
\end{equation}
From this plots it is clear that the density signal, i.e. an unique feature of our estimator $\DL$, has an SNR$\gtrsim 1 $ at low redshift. However it should be pointed out that this does not immediately translate to the possibility of detection since one would also have to deal with the separation of the different components of the four contributions, which is beyond the scope of this paper.
In the case of ET+2CE, the SNR of the angular power spectrum of $\Delta\ln\DL$ is smaller than one in all the redshift bins when we consider BBH, while thanks to the large number of BNS the signal could be detected at low redshifts.  On the other hand, in the BBO/DECIGO case, the large number of detected BNS combined with a larger angular resolution, makes possible to detect with high significance the density contribution in the BNS case. Therefore it will give the possibility to obtain information on the cosmological parameters and the bias of the GWs that affect the GR corrections by using the observable introduced in this work. If instead we consider BBH in the BBO/DECIGO case it will not be possible to claim a strong detection given the large shot noise, which is not compensated by the more precise measurements (both in LD and in angular resolution).

\section{Fitting function}
\label{app:fitting_formula}
To use the error estimates we find in this work, we provide a few fitting formulas for the errors associated to $\DL$ due to the GR corrections. 
We fit all the curves with the function
\begin{equation}
    \label{eq:fit}
    \frac{\sigma_{\rm anis}}{\D_L} = A + Bz + Cz^2 + Dz^3 + Ez^{-1}\,,
\end{equation}
while in the following tables we report the values divided in the non-weighted case (the one treated in \cite{Bertacca:2017vod}) and the weighted case. In Table \ref{table_ETCE} we show the numerical coefficients for ET+2CE, while in Table \ref{table_BBO} we show the results for BBO/DECIGO. Moreover we report the fits for both BBH and BNS and we also note that for BBO/DECIGO in the non-weighted case there are no differences between BBH and BNS. This happens because the main driver of difference between the two is the angular precision, that reaches non-linear scales in both case, above which our analysis stops.

\begin{table}[H]
\begin{center}
\begin{tabular}{|l|l|l|l|l|l|l|}
\hline
 \multicolumn{2}{|l|}{ET+2CE}& A &  B &  C & D & E \\ \hline
\multirow{2}{4em}{non -weighted} & BBH &-0.00127& 0.00442& -0.0013& 0.00014& 0.00089\\ \cline{2-7}
 &BNS  &-0.00115& 0.00252& -0.00065& $6 \times 10^{-5}$& 0.00088 \\ \hline
\multirow{2}{4em}{weighted} &BBH& 0.00534& -0.00361& 0.00181& -0.00024& 0.00126  \\ \cline{2-7} 
 & BNS &0.0083& -0.00805& 0.00301& -0.00036& 0.00179\\ \hline

\end{tabular}
\end{center}
 \caption{Coefficients of Eq. \eqref{eq:fit} to fit the error estimates plotted in Figure \ref{fig:fitting_plot} for the case of ET+2CE.}\label{table_ETCE}
 \end{table}



\begin{table}[H]
\begin{center}
\begin{tabular}{|l|l|l|l|l|l|l|}
\hline
 \multicolumn{2}{|l|}{BBO/DECIGO}& A &  B &  C & D & E \\ \hline 
\multirow{2}{4em}{non-weighted} & BBH &\multirow{2}{4em}{0.00383}&\multirow{2}{4em}{0.00924}& \multirow{2}{4em}{-0.00054}&\multirow{2}{4em}{ $-6\times 10^{-5}$}& \multirow{2}{4em}{0.00112}\\ \cline{2-2}
 &BNS & &&&& \\ \hline
\multirow{2}{4em}{weighted} &BBH& 0.00327& 0.00391& 0.00115& -0.00026& 0.00139  \\ \cline{2-7} 
 & BNS &0.01149& 0.00996& -0.00296& 0.00023& 0.00131  \\ \hline

\end{tabular}
\end{center}
  \caption{In this table we report the coefficients of Eq.\eqref{eq:fit} to fit the error estimates plotted in Fig.\ref{fig:fitting_plot} for the case of BBO/DECIGO.}\label{table_BBO}
\end{table}

\bibliographystyle{JHEP}
\bibliography{bibliography.bib}

\end{document}